\documentclass{aa}  
\usepackage{graphicx}
\usepackage{txfonts}
 \usepackage{aalongtable,lscape}
 \usepackage{colortbl}
 \usepackage{natbib}
\usepackage{url}
 \bibpunct{(}{)}{;}{a}{}{,}
\begin{document}
   \title{Morphological evolution of $z\sim1$ galaxies from deep K-band AO imaging in the COSMOS deep field \thanks{Based on ESO observations at the VLT. Programmes P73.A-0814A and P75.A-0569A and 175.A-0839 (zCOSMOS).} \fnmsep \thanks{Based on observations obtained with MegaPrime/MegaCam, a joint project of the CFHT and CEA/DAPNIA, at the Canada-France-Hawaii Telescope (CFHT) which is operated by the National Research Council (NRC) of Canada, the Institut National des Science de l'Univers of the Centre National de la Recherche Scientifique (CNRS) of France, and the University of Hawaii. This work is based in part on data products produced at TERAPIX and the Canadian Astronomy Data Centre as part of the Canada-France-Hawaii Telescope Legacy Survey, a collaborative project of the NRC and CNRS.}
   }

   %\subtitle{I. Overviewing the $\kappa$-mechanism}
\author{M. Huertas-Company
          \inst{1,4},
          D. Rouan \inst{1},
          G. Soucail \inst{2}, 
          O. Le F\`evre \inst{3},
          L. Tasca \inst{3},
          T. Contini \inst{2}
          }

   \institute{LESIA-Paris Observatory, 
   5 Place Jules Janssen, 92195 Meudon, France\\
              \email{marc.huertas@obspm.fr}         
            \and
Laboratoire d'Astrophysique de Toulouse-Tarbes, CNRS-UMR 5572 and Universit\'e Paul Sabatier Toulouse III, 14 Avenue Belin, 31400 Toulouse,  France
            \and
           LAM-Marseille Observatory, Traverse du Siphon-Les trois Lucs BP8-13376 Marseille Cedex 12, France
           \and
           IAA-C/ Camino Bajo de Hu\'etor, 50 - 18008 Granada, Spain}

   \date{Received <date> / Accepted <date>}

% \abstract{}{}{}{}{} 
% 5 {} token are mandatory
 
  \abstract
   {We present the results of an imaging program of
   distant galaxies ($z\sim0.8$) at high spatial resolution ($\sim0.1"$)
   aiming at studying their morphological evolution. We observed 7
   fields of $1'\times1'$ with the NACO Adaptive Optics system (VLT)
   in Ks ($2.16\mu m$) band with typical $V\sim14$ guide stars and 3h
   integration time per field. Observed fields are selected within
   the COSMOS survey area, in which multi-wavelength photometric and
   spectroscopic observations are ongoing. High angular-resolution K-band data %is less dependent on the recent history of star formation, which peaks in the UV in rest frame, and give thus a more objective view of galaxy evolution than optical observations. 
   have the advantage of probing old stellar 
populations in the rest-frame, enabling a determination of galaxy morphological 
types unaffected by recent star formation, which are more closely linked to the 
underlying mass than classical optical morphology studies (HST).
   Adaptive optics on ground based telescopes is the only method today for obtaining
   such a high resolution in the K-band, but it suffers from limitations since only small fields are observable and long integration times are necessary. }
   {In this paper we show that reliable results
   can be obtained and establish a first basis for larger observing
   programs.}
   {We analyze the morphologies by means of B/D (bulge/disk) decomposition
   with GIM2D and C-A (concentration-asymmetry) estimators for $79$
   galaxies with magnitudes between $K_s=17-23$ and classify
   them into three main morphological types (late type, early type and
   irregulars). Automated and objective classification allows precise
   error estimation. Simulations and comparisons with seeing-limited
   (CFHT/Megacam) and space (HST/ACS) data are carried out to
   evaluate the accuracy of adaptive optics-based observations for
   morphological purposes.  }
   {We obtain the first estimate of the distribution
   of galaxy types at redshift $z\sim1$ as measured from the near
   infrared at high spatial resolution. We show that galaxy parameters
   (disk scale length, bulge effective radius, and bulge fraction) can
   be estimated with a random error lower than $20\%$ for the bulge
   fraction up to $K_s=19$ ($AB=21$) and that classification into the
   three main morphological types can be done up to $K_s=20$ ($AB=22$)
   with at least 70\% of the identifications correct. We used the known photometric redshifts to obtain a redshift distribution over 2 redshift bins ($z<0.8$, $0.8<z<1.5$) for each morphological type. 
   %We find an excess
   %of irregular objects ($\sim30\%$), while no evolution indices seem
   %to appear in the elliptical ($\sim20\%$) and spiral ($\sim50\%$)
   %populations. The need for future larger programs is emphasized.
    }
   {}


\authorrunning{Huertas-Company et al}
\titlerunning{Morphology of $z\sim1$ galaxies with adaptive optics}

   \maketitle
%
%________________________________________________________________

\section{Introduction}

The process of galaxy formation and the way galaxies evolve is still one
of the key unresolved problems in modern astrophysics.  In the
currently accepted hierarchical picture of structure formation, galaxies
are thought to be embedded in massive dark halos that grow from density
fluctuations in the early universe \citep{Fa80} and initially contain
baryons in a hot gaseous phase. This gas subsequently cools, and some
fraction eventually condenses into stars. \citep{Lil96,Mad98}. However,
many of the physical details remain uncertain, in particular the process
and history of mass assembly. One classical observational way to test
those models is to classify galaxies according to morphological  criteria
defined in the nearby Universe \citep{Hub36,deVauc48,San61}, which can be
related to physical properties, and to follow this classification across
time. \citep{Ab96,Sim02,Abr03}. Indeed, since galaxies were recognized as
distinct physical systems, one of the main goals in extragalactic astronomy
has been to characterize their shapes. Comparison of distant populations with
the ones found in the nearby Universe might help to clarify the role
of merging as one of the main drivers in galaxy evolution.
\citep{Co00, Ba96}. Progress in this field has
come from observing deeper and larger samples, but also from obtaining
higher spatial resolution at a given flux and at a given redshift.  In the
visible, progress has been simultaneous on those two fronts, thanks in particular to the
ultra-deep HDF fields observed with the Hubble Space Telescope. HST imaging has brought observational evidence that galaxy evolution
is differentiated with respect to morphological type and that a large
fraction of distant galaxies have peculiar morphologies that do not fit
into the elliptical-spiral Hubble sequence. \citep{Bri98, Wo03, Il06}. These results can however be biased by the fact that most of the sampled
galaxies are at large redshift and are analyzed through their UV
rest-frame emission, which is more sensitive to star formation processes
and to extinction. Moreover,
it  now seems clear that evolution strongly depends on the galaxy mass
in the sense that massive systems appear to have star formation histories that peak at higher redshifts,  whereas less massive systems have star formation histories that peak at progressively lower redshifts and are extended over a longer time interval (downsizing scenario; \citealp{Cow96, Bri00, Bun05}). This could explain the fact that the population of massive E/S0 seems to be in place at  $z\sim1$ and evolve passively towards lower redshifts \citep{DeLu06, Zucca06}. However, most of these studies are based on spectral type classifications, and their interpretation in the framework of galaxy formation is not straightforward since galaxies move from one spectral class to another by passive evolution of the stellar populations. \\

In this context, high-resolution near-infrared observations are particularly important
because the K-band flux is less dependent on the recent history of star formation, which peaks in the UV in rest frame, and thus gives a galaxy type from the distribution of old stars that is more closely related to the underlying total mass than optical observations. This is why a large number of K-band surveys have been carried out using ground-based telescopes with different spatial coverages and limiting magnitudes in order to perform cosmological tests by means of galaxy counts essentially \citep{Gard93,
Glaze95,Mcl95,Be98,Mc00,Ma01}. However, no morphological information can
be found in particular due to the seeing limited resolution, even with superb image quality as in the ongoing CFHT/WIRCAM survey. 
%Indeed, resolution
%is crucial in order to probe the numerous small objects present in a
%hierarchical Universe. 
\cite{Con00} prove that only when the ratio
of the angular diameter substending $0.5h^{-1}_{75}kpc$ at a given
distance to the angular resolution of the image is around $1$ can reliable
morphological estimators such as the asymmetry be obtained. \\

Consequently, adaptive optics (AO) installed on ground-based telescopes will be a powerful method for obtaining near-infrared high-resolution data in the future and an excellent complement to space observations, such as those that will be performed with the HST (WFC3, 2008) and with the JWST (2014).  

%since no space observations are programmed until 2014 with JWST. 
%\textbf{REMOVED SENTENCE}

The use of classical AO for deep surveys suffers, however, from inherent limitations such as the non-stationary PSF on long integration times and the finite isoplanetic field.  This is why preliminary studies to probe the accuracy of adaptive
optics are required, before launching very wide surveys.  In particular,
it is important to determine whether automated morphological classifications
can be performed. Indeed, given the large number of galaxies, such automated
methods for morphological classifications are desirable. \\

\cite{Mino05} published first results based on adaptive optics
 observations. They achieved the impressive limiting magnitude of $K\sim 24.7$
with the Subaru Adaptive Optics system with a total integration time of
$26.8 hr$ over one single field of $1'\times1'$. They proved that the use
of adaptive optics significantly improves detection of faint sources but did not obtain morphological information. In a recently submitted paper, \cite{Cre06} perform a morphological analysis based on AO data for the first time. They observed a $15$ $arcmin^2$ in the $K_s$ band with NACO (SWAN survey) and classified distant galaxies into two morphological bins (late type, early type) by performing a model fitting with a Sersic law. They compared number counts and size-magnitude relation, for early and late-types separately, with hierarchical and pure luminosity evolution (PLE) models, respectively. They conclude that hierarchical models are not consistent with the observed number counts of elliptical galaxies and that PLE models are preferred.  However, as discussed in several studies \citep{Gard98}, despite galaxy counts still being one of the classical cosmological tests, their
interpretation remains difficult. It is  thus not realistic to expect galaxy counts alone to strongly constrain the cosmological geometry or even to constrain galaxy evolution.  A more complete study needs redshift estimates, which is lacking in the SWAN survey. That is the main reason for having select the COSMOS \citep{Sco05} field to conduct our pilot program  since multi-wavelength photometric and spectroscopic observations are performed. This ensures a reliable redshift estimate for all our objects. \\

%ZZZ Presenter SWAN !!! ZZZ

  %Since the first
%Hubble's subjective and visual classifications, new methods of classifying
%galaxies have been proposed. These classification methods can be either
%parametric (model-based) or non-parametric. Non-parametric classifiers
%include the C-A system \citep{Ab94,Ab96,Con00} and artificial neural
%networks trained from visual classification \citep{Bu92}. Parametric
%classifiers include bulge+disk decomposition \citep{Sch95,Sim02}.

In this paper, we continue this AO validation task by morphologically classifying a sample of $79$ galaxies
observed %with the NAOS/CONICA adaptive optics system distributed over $7arcmin^2$ area 
using parametric
(GIM2D, \cite{Sim02}) and non-parametric (C-A, \cite{Ab94}) methods and comparing them. Fields, observed with NAOS/CONICA adaptive optics system, are distributed over a $7arcmin^2$ area. We obtain for the first time an estimate of the
distribution of galaxies in three morphological types (E/S0,S,Irr) at redshift $z\sim1$ as measured from the
near infrared at high spatial resolution. We then use the photometric
redshifts to look for evolution clues as a function of morphology. \\

The paper proceeds as follows: the data set and the reduction procedures
are presented in the next section. In \S~\ref{sec:detec}, we focus on the detection
procedures and the sample completeness. In \S~\ref{sec:zphot} we discuss the estimate
of redshifts using the multi-wavelength photometric data from COSMOS. The morphological analysis is
discussed in \S~\ref{sec:morpho} using bulge/disk decompositions and concentration
and asymmetry estimators. Simulations
for error characterization are carried out for both classification
methods, and comparisons between classifications are shown. In \S~\ref{sec:comparisons}
we compare the data with ground and space observations and use
those comparisons to discus the morphological evolution in the last section. Throughout this
paper magnitudes are given in the Vega system. We use the following
cosmological parameters: $H_0 = 70\,\mathrm{km\,s^{-1}\,Mpc^{-1}}$ and
$(\Omega_\mathrm{M}, \Omega_\Lambda, \Omega_\mathrm{k})=(0.3,0.7,0.0)$.

 \section{The data set}
 \label{sec:dataset}
\begin{figure}
 \resizebox{\hsize}{!}{\includegraphics{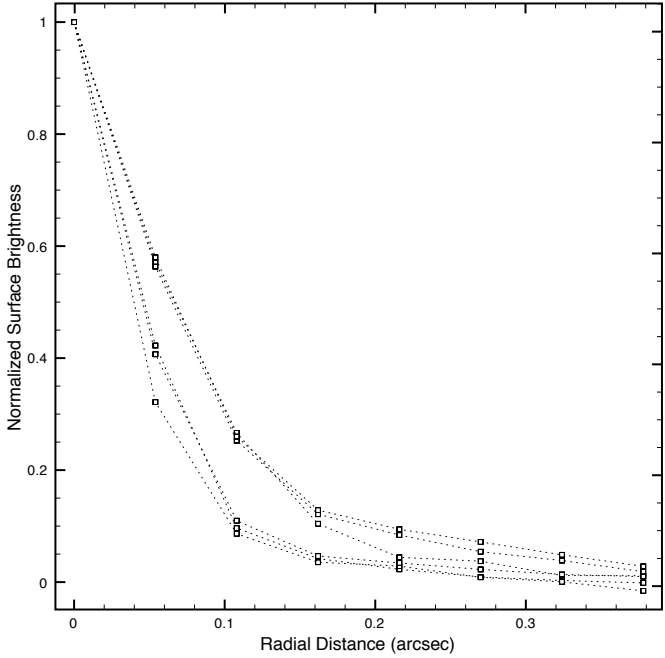}}
 \caption{Radial
 profile of all detected stellar objects with $K_s<19$. The mean resolution is $\sim
 0.1\arcsec$, broader than the telescope diffraction limit, as a
 consequence of the long integration time ($\sim 3h$) and the large fields.}
 \label{fig:psf_profile}
 \end{figure}

Seven fields of $1\arcmin \times 1\arcmin$ were observed in $K_s$ band ($2.16 \mu m$) with the
NAOS/CONICA-assisted infrared camera installed on the VLT. \footnote{P73.A-0814A and
P75.A-0569A}. The fields were selected within the COSMOS survey
area\footnote{\url{http://www.astro.caltech.edu/cosmos/}}. In order
to ensure a reliable AO correction, relatively bright stars ($V\sim
14$) were selected. We added a color criterium ($B-R < 1.0$) in order
to benefit from a large attenuation of the flux in the near-IR and thus a smaller occultation of the central region of the images. The pixel
scale ($0.054\arcsec$)  was chosen to be twice the Nyquist-Shannon requirements with respect to the telescope diffraction limit in order to have larger fields. With such a pixel sampling, the 
PSF FWHM would only be one pixel width in the limit of perfect AO correction. However, we will 
 show in Sect. 5 that our PSF reconstruction procedure can handle such undersampled data.
 We also note that only partial AO correction was actually achieved during our observations,
 and our PSF reconstruction is thus more than adequate here. Figure~\ref{fig:psf_profile} compares the radial profiles of 5 detected stellar objects. Indeed, this program is pushed to its limits in terms of field size, exposure time, and brightness of the guide star.

 \begin{table} \centering 
 \begin{tabular}{c c c c c} \hline \hline \noalign{\smallskip} 
Field & $\alpha$ & $\delta$ & exp.time (s) & Seeing (arcsec)\\ 
\noalign{\smallskip} \hline \noalign{\smallskip}
  STAR1 & 10:00:16 & +02:16:22 & 10350 & 0.09 \\
    STAR2 & 10:00:52 & +02:19:52 & 7650 & \_ \\
      STAR3 & 10:00:10 & +02:06:08 & 7650 & 0.12 \\
        STAR4 & 09:59:52 & +02:05:00 & 7170 & 0.08 \\
  STAR5 & 10:00:14 & +02:09:09 & 10200 & 0.09 \\
    STAR6 & 10:00:02 & +02:06:57 & 7650 & 0.13\\
      STAR10 & 09:59:56 & +02:04:07 & 10000 & \_\\        
     
\noalign{\smallskip}\hline
\end{tabular}
  \caption{Summary of observations for the seven analyzed fields. The mean seeing is estimated when faint stars were detected.}
  \label{tbl:obs_sum}
\end{table}

Data are reduced in a standard way: exposures are taken in the
auto-jitter mode, which means that the pointing is randomly shifted within
a 7\arcsec box, in order to improve flat fielding, bad pixel correction, and sky background withdrawal. The sky value in each pixel is estimated by performing a clipped median of all the exposure frames: the $10\%$ faintest and brightest values are removed from the computation.
Cosmic ray and flat corrections were applied,
and recentering was done using a standard cross-correlation method. Recentering is done at a sub-pixel level. For that purpose a cubic interpolation of resampling was performed. After stacking, a global estimate of the sky background was performed by computing the stacked image spatial median. The final image was obtained after substraction of this value. 

 %ZZZ a detailler un peu, en particulier la soustraction du ciel? ZZZ.

Photometric zeropoints were determined using 2MASS stars
\citep{Klein92}. We performed aperture photometry on the guide stars and compared it to 2MASS data to deduce the zeropoints. Note that the change of detector between periods
P73 and P75 resulted in different zeropoint values for each run. The average zeropoint
for the first period is: $22.82\pm 0.06 $ and for the second period
$23.29\pm 0.06$. We also used the ESO calibration data set standard stars \citep{Per98} to validate our measurements (ESO pipeline computations and our own measurements on the standard stars images). There is good agreement between all these values.

\section{Detection and completeness}
\label{sec:detec}
\begin{figure} 
 \centering 
\resizebox{\hsize}{!}{\includegraphics{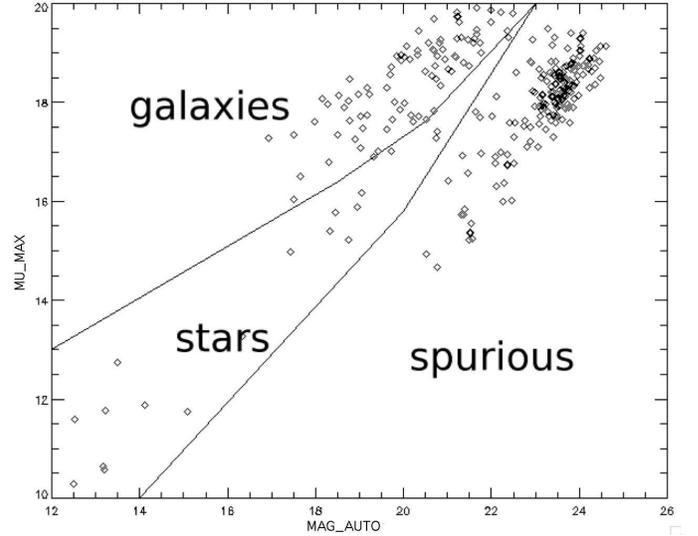}}
 \caption{Objects classification using the MAG\_AUTO-MU\_MAX plane. }
 %objects with high peak surface brightness and low global brightness
 %are considered as false detections.} 
 \label{fig:gal_sep} 
 \end{figure}

All objects having a $3\sigma$ signal above sky, over $4$ four contiguous pixels
are detected using \textsc{SExtractor} \citep{Ber96}. 
We decided to apply this low
detection threshold, even if the main goal of this paper is to perform a morphological analysis, for two main reasons. First we want to test the ability of adaptive optics based observations to obtain morphology, so we are seeking the limits; and second, we wanted to be sure that no objects are lost when computing number counts to compare with other near-infrared observations (see Appendix A). 
We find $285$ objects over the $7$ fields. We then performed a
cleaning task in order to separate galaxies from stars and spurious
detections. This was made using the \textsc{SExtractor} MU\_MAX and
MAG\_AUTO parameters that give the peak surface brightness above the background
and the Kron-like elliptical aperture magnitude, respectively. The distribution
of objects in this parameter space clearly defines three regions that separate extended sources from point-like or non-resolved
sources and from spurious detections (Fig.~\ref{fig:gal_sep}). In this
separation scheme, objects with very faint magnitude and high peak
surface brightness are considered as false detections. Boundaries were drawn
manually and a visual inspection confirms that known stars in the field
are indeed identified as point sources.  We consequently identified $79$
galaxies, $19$ stars (or unresolved objects), and $187$ spurious objects
in the whole field.

 \begin{figure*}
 \centering 
 \includegraphics[width=12cm]{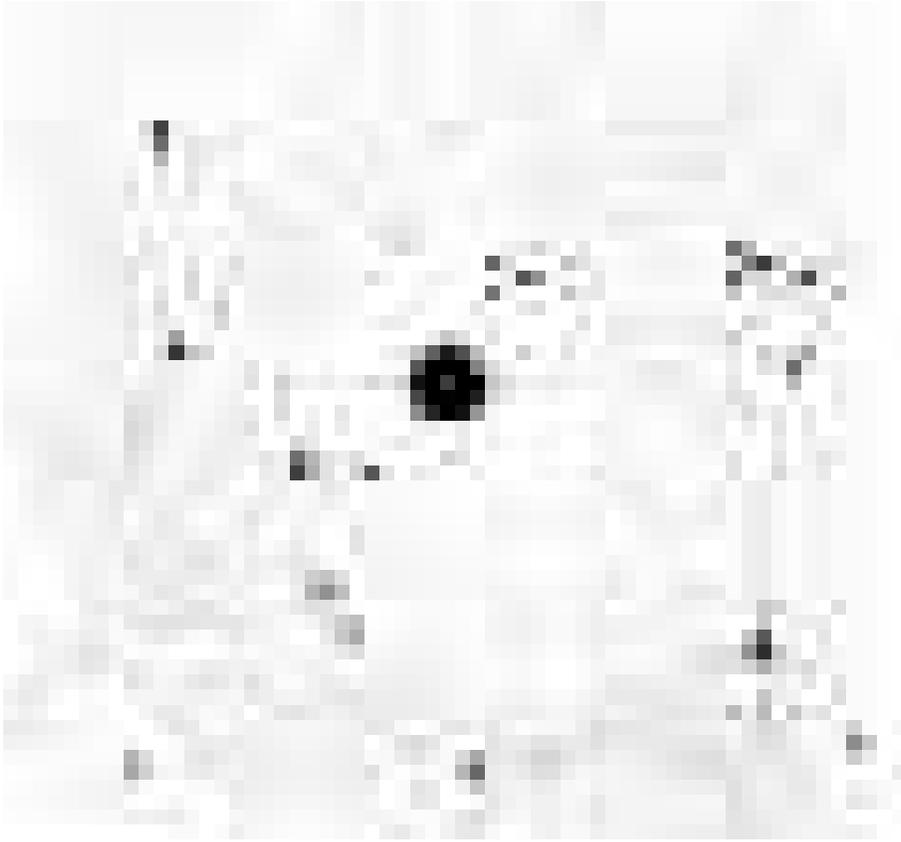}
 \caption{NAOS/CONICA $K_s$-band image of the field centered at
 $\alpha=$ 10:00:16, $\delta=$ +02:16:22 . The total integration time
 is 10350 s. The field size is $1\arcmin \times1\arcmin$ with a pixel scale of
 54 mas. Circles are detected galaxies and boxes are stars. The stellar
 FWHM is measured to be $0.1\arcsec$. The bright star at the center of the
 image is used as the AO guide star.  } 
 \label{fig:field} 
 \end{figure*}

\begin{figure}
 \centering 
\resizebox{\hsize}{!}{\includegraphics{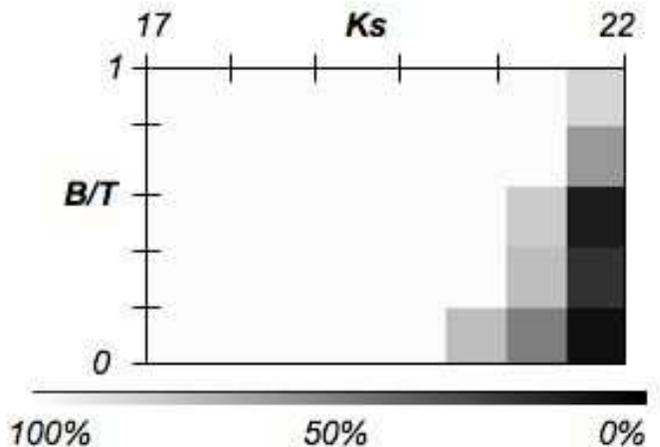}}
 \caption{Completeness for extended sources. Galaxies with parameters ($r_d$,$r_e$,$B/T$) uniformly distributed are simulated
 and placed at random positions in the fields.}
 \label{fig:compgal} 
 \end{figure}

The sample completeness for point sources was estimated by creating
artificial point sources from fields stars (see sect. 5 for detailed
explanations) with apparent magnitudes ranging from $K_s=18$ to $K_s=24$
and placing them at random positions. We ran \textsc{SExtractor} with the same
configuration as for real sources and looked for the fraction of detected
objects. We find that the sample is $50\%$ complete
at $K_s=22.5$ (or $AB=24.5$) for point sources. Completeness for extended
sources is estimated in a similar way: we generate $1000$ galaxies
with exponential and de Vaucouleurs profiles of different morphological
types (bulge fraction ranging from 0 to 1) and with galaxy parameters
uniformly distributed. In particular, the sizes of disks and bulges are distributed uniformly between $0^{"}<r_d<0.7^{"}$ and $0^{"}<r_e<0.7^{"}$ as detailed in Sect. \S~\ref{sec:morpho}. This leads to half-luminosity radii ranging from $0^{"}<r_h<1^{"}$. We find that the sample is 50\% complete at
$K_s=21.5$ (or $AB=23.5$) for this population of extended sources (Fig.~\ref{fig:compgal}). We used this completeness to compute number counts and to compare it to other near-infrared surveys in appendix~\ref{sec:app}.

%  \begin{figure}
% \centering \includegraphics[scale=0.3]{completude.eps}
% \caption{Completeness for point sources. Artificial point sources
% were generated and embedded at random positions in the fields. }
% \label{fig:comp} \end{figure}

\section{Photometric redshifts}
\label{sec:zphot}
Galaxy number counts provide a useful
    description of galaxy populations but suffer from
    numerous degeneracies when trying to trace the
    evolution of galaxy populations. Model predictions are subject to uncertainties in the spectral
energy distributions and evolution of galaxies and in the free parameters
specifying the luminosity function, the cosmological geometry, the number
and distribution of galaxy types, and the effect of dust and merging.
The need to have redshift information is
therefore the reason for selecting the NACO fields within the ongoing
Cosmic Evolution Survey (COSMOS) in which multi-$\lambda$ and spectroscopic observations
are performed. COSMOS is designed to probe the correlated evolution of galaxies, star formation, active galactic nuclei (AGN) and dark matter (DM) with large-scale structure (LSS) over the redshift range z = 0.5 to 3. The survey includes multi-wavelength imaging and spectroscopy from X-ray to radio wavelengths covering a 2 square deg area, including HST imaging of the entire field.
% In addition, these fields are observed with the CFHT in
%5 photometric bands $(u*,g',r',i',z')$ from the near-UV to visible domain.

 \begin{figure} 
 \centering 
  \resizebox{\hsize}{!}{\includegraphics{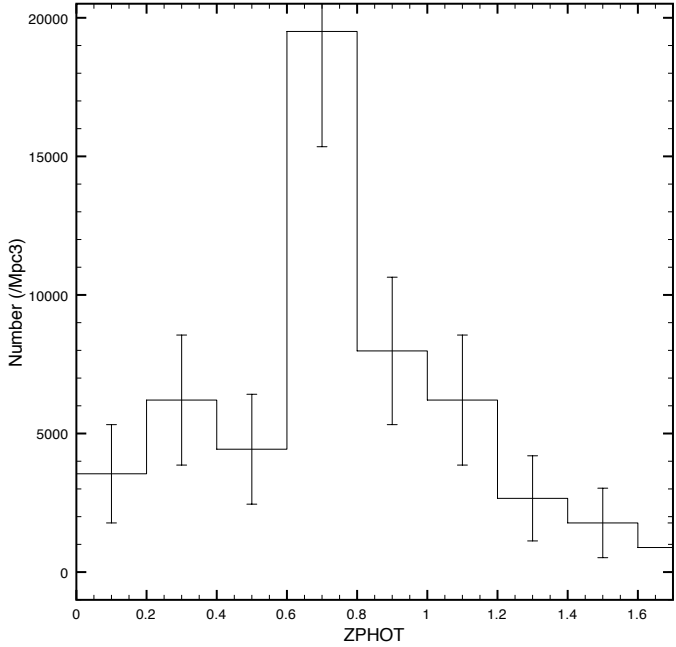}}
 \caption{  Le Phare photometric redshift distribution for the 60 matched
 objects. The distribution is peaked around $z\sim0.8$,  in good
 agreement with the predictions of simple PLE models. Error bars show
 poissonnian errors.} 
 \label{fig:histo_z} 
 \end{figure}

% \subsection{Le Phare photometric redshifts}

All these data are used for a direct estimate of the
photometric redshifts of the galaxies detected in the NACO fields,
computed with the code {\it{Le
Phare}}\footnote{\url{http://www.lam.oamp.fr/arnouts/LE_PHARE.html}}.  
A standard $\chi^2$ method
is implemented, including an iterative zero-point
refinement combined with a template optimization procedure and the
application of a Bayesian approach \citep{Ilb06}. We used 1095
spectroscopic redshifts taken from the zCOSMOS Survey \citep{Lilly06} to measure the photometric redshifts. 
This method allows to
reach an accuracy of $\sigma_{\Delta z}/ (1+z_s) = 0.031$ with $1.0 \%$ catastrophic errors, defined as $\Delta z /(1+z_s) >0.15$.

%ZZZ preciser la reference Capack et al. ZZZ

%ZZZ preciser la reference Capack et al. ZZZ

The multi-color catalog of the COSMOS survey (Capack et al. 2006) consists
of photometry measurements over 3 arcsecond diameter apertures for deep
$B_j, V_j, g+, r+, i+, z+$ Subaru data taken with SuprimeCam, $u*, i*$
bands with MegaCam (CFHT), $u',g',r',i',z'$ information from the Sloan
Digital Sky Survey (SDSS), $K_s$ magnitude from KPNO/CTIO, and $F816W$
HST/ACS magnitude. Objects were matched between the COSMOS and the NACO
catalogs within a radius of 2\arcsec\ which takes possible
astrometry differences between the calalogues into account.  We matched 60 objects out
of the 79 detected in the NACO fields.  Fig.~\ref{fig:histo_z} shows
the photometric redshift distribution for these 60 matched objects.
As expected for a galaxy sample limited to $K_s=22$, the redshift 
distribution peaks around $z \sim 0.8$ \citep{Mig05}.

%A subset of bright NACO galaxies is used to show in appendix A the
%improvement in the photoz measurement due to the use of the deeper
%NACO K-band

 \section{Automated morphology classifications}
 \label{sec:morpho}
The 79 objects identified as extended sources are morphologically
classified using two automated methods based on direct model fitting and
on learning classification. Automated classifications have
the fundamental properties of being objective, thus reproducible, and
they allow a precise error characterization. %Moreover, in the
%present epoch of very large and deep surveys visual classifications appear
%to be outdated because of the huge amount of data. 
We proceeded in two
steps: first we detected irregular objects using asymmetry estimators,
 and then we separated the regular objects between early and late type
objects. \\

Throughout this section, we use extensive simulations for error
estimates and calibration of the automated classifications as explained below. For all the simulations, we assumed that bulges
are pure de Vaucouleurs profiles (n = 4) and that disks are
exponential profiles. We then generate galaxies with parameters uniformly distributed in the following parameter space: $0<B/T<1$, $0^{"}<r_d<0.7^{"}$, $0^{"}<r_e<0.7^{"}$, $0^{\circ}<i<70^{\circ}$, and $0<e<0.7$. Both bulge and disk position angles were fixed to $90^{\circ}$.
The goal of these simulations is to characterize biases and
errors; the uniformity of the parameter distributions adopted
here is therefore perfectly suitable, even though real galaxy parameters
may not be so distributed. For the same reason, we
do not take any redshift effect into account. Each simulation
was convolved with the reconstructed PSF as explained in~\ref{sec:gim2d}.
The same PSF was used in both creating and analyzing the simulations,
so the results will not include any error due to PSF
mismatch. In order to simulate background noise, objects are
embedded at random positions in the fields and detected with
the same \textsc{SExtractor} parameters as for the real sources.

%All along this section, we use extensive simulations for error estimates and calibration of the automated classifications as detailed further. For all the simulations, we assume that bulges are pure de Vaucouleurs profiles and that disks are exponential profiles. We then generate galaxies with parameters uniformly distributed in the following parameter space: $0<B/T<1$, $0^{"}<r_d<0.7^{"}$, $0^{"}<r_e<0.7^{"}$, $0^{\circ}<i<70^{\circ}$ and $0<e<0.7$. The position angles are set to 0. The models are then convolved  with a PSF computed as explained in~\ref{sec:gim2d} and dropped in the observed fields to reproduce the observing conditions. The same detection procedures as for real galaxies are employed.

 \subsection{Irregular objects}
% \subsubsection{Detection procedure}
\label{sec:irr}
The detection of objects presenting irregularities is made using the
concentration and asymmetry estimators \citep{Ab94,Ab96}. Concentration is computed as the ratio of the flux between the inner and outer isophotes of normalized radii 0.3 and 1 within the isophotal area enclosed by pixels $3\sigma$ above the sky level. The corresponding limiting surface brightness varies between $\mu=18.76-20.39$ $mag.arcsec^{-2}$ because of the variations in the exposure time between the different fields and the intrinsic variation of the sky level, which is important for infrared ground-based observations. Indeed, the value of C is quite sensitive to the estimate of the background level since an error in this value will result in different limiting isophotes, and a fraction of the galaxy flux can be lost. To estimate the error in C introduced by the different isophote levels, we computed the variations in the C value for variations in the limiting surface brightness of $\Delta \mu=1$. We found a fairly small error ($\Delta C \sim 0.06$), so we decided not to apply any corrections. Asymmetry
(A) was obtained by rotating the galaxy image about its center by $180^{¡}$ and
self-substracting it to the unrotated image after sky substraction. Local sky level is estimated using \textsc{SExtractor} output parameter \emph{BACKGROUND} that gives the background level at the galaxy centroid position.
The center of rotation was determined by first smoothing the galaxy image with a Gaussian kernel of $\sigma =1$ and then choosing the location of the maximum pixel as the center, as explained in \cite{Ab96}.
%The galaxy center is
%estimated \textbf{by finding the center that minimizes the rotational asymmetry, as explained in \cite{Con00}. We look for the center in a (-3,+3) pixel area, around the initial guess, which is the object barycenter as computed by \textsc{SExtractor}}. 
Since the
absolute value for the residual light is used, noise in the images
shows up as a small positive A signal even, in perfectly symmetrical
objects.  \citep{Con00}. This is why we applied a noise correction to
the computation: the value of A in a portion of sky with area equal to
that enclosed by the galaxy isophote.The definition of A that
is used in the rest of the paper is:

 \begin{equation} 
A = \frac{1}{2}\left(\frac{\sum
|I(i,j)-I_{180}(i,j)|}{\sum I(i,j)} - \frac{\sum
 |B(i,j)-B_{180}(i,j)|}{\sum I(i,j)}\right) \label{eq:Asym} 
 \end{equation}

%\textbf{The value of A depends on the  }

%ZZZ Rappeler la definition du parametre C ZZZ.

\begin{figure*}     
\centering          
\includegraphics[width=0.48\textwidth,height=0.4\textwidth]{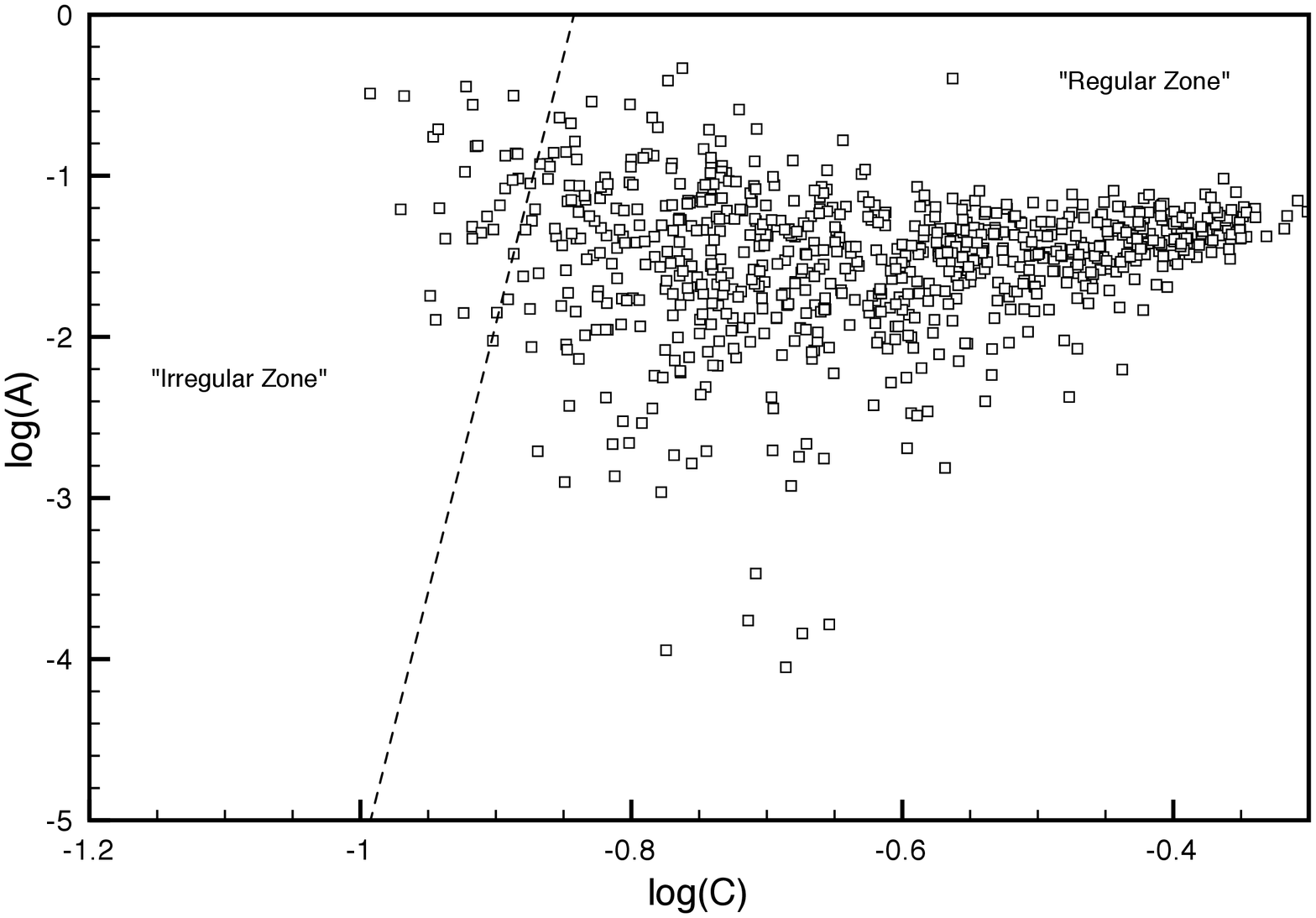} 
\includegraphics[width=0.48\textwidth,height=0.4\textwidth]{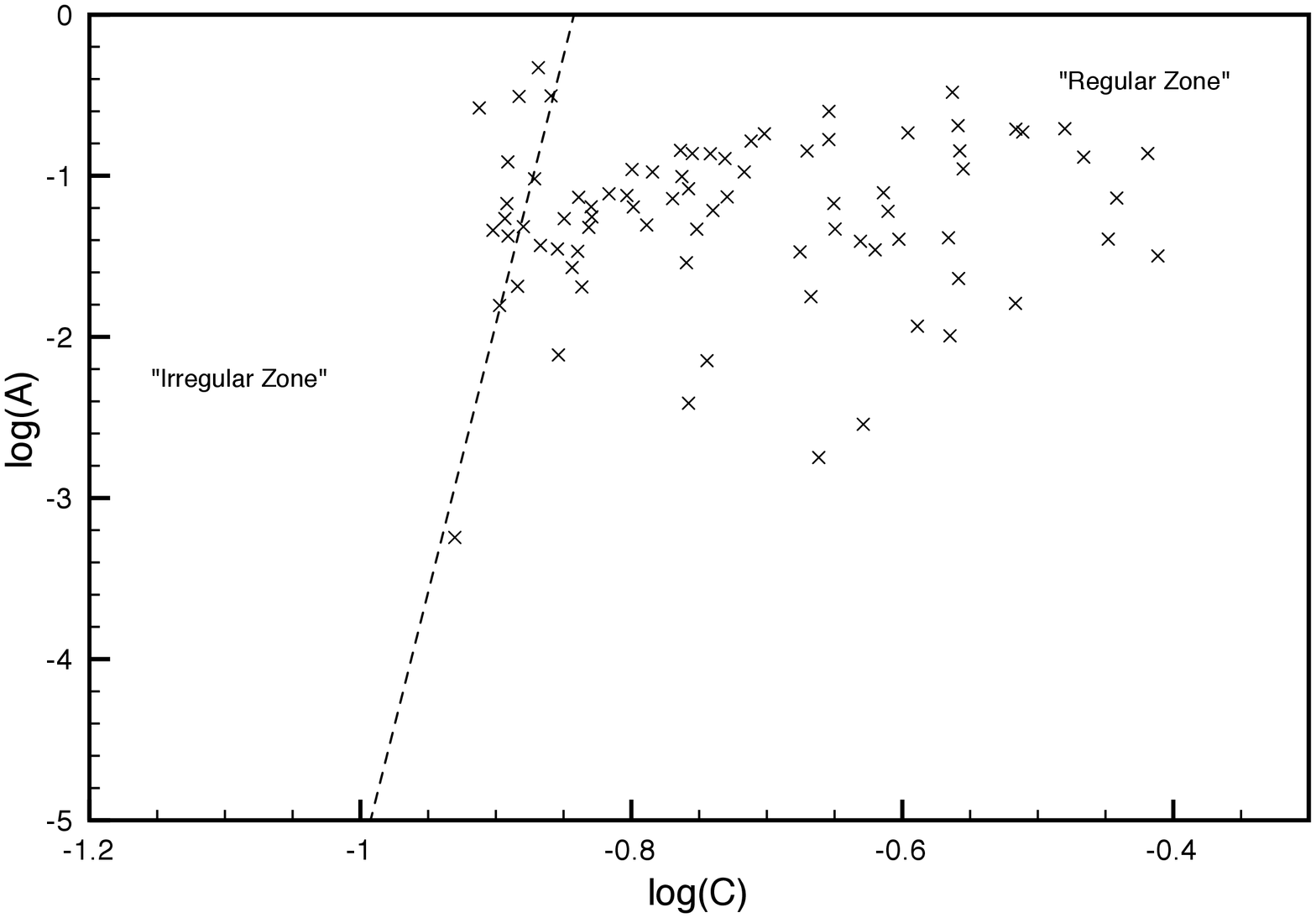}
\caption{Separation between regular and irregular objects. Left: simulated objects (empty squares),
Right: real objects (crosses)
%ZZZ Plot de droite a refaire sans les "empty squares" ZZZ
.}
\label{fig:irr} \end{figure*}

To establish the boundaries between regular and irregular objects,
a calibration of the C-A plane is needed. We thus simulated a set of
galaxies with different galaxy parameters and magnitudes ranging between
$17<K_s<23$ that we embedded in the real images. We computed the C and
A parameters of these objects and plotted  the C-A plane. Since
irregular objects cannot be simulated in a meaningful way, a kind of
extrapolation was employed, based on two facts: a)irregular galaxies have flatter photometric profiles (less concentrated)
and to be more asymmetric than regular objects; b)those objects are not
present in the simulated sample. We thus defined the irregular zone as
the upper left corner of the C-A plane where no simulated objects are
found (Fig.~\ref{fig:irr}).

 \begin{figure}
 \centering
   \resizebox{\hsize}{!}{\includegraphics{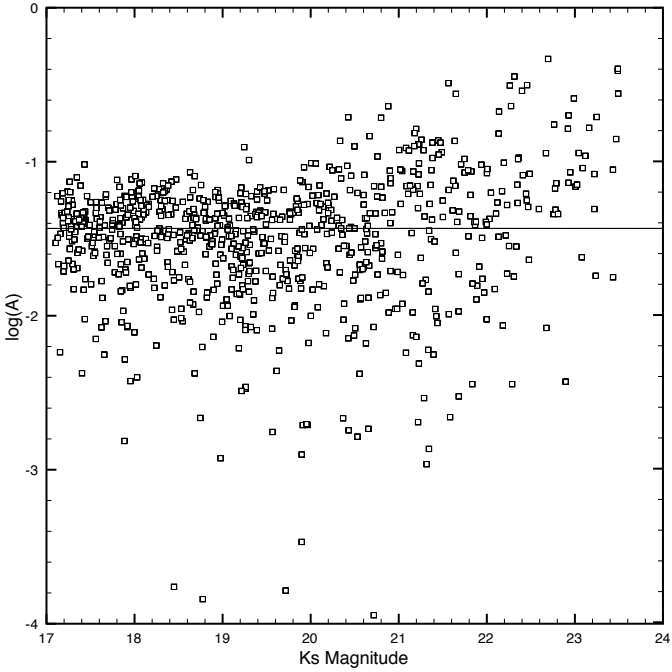}}
 \caption{Asymmetry versus magnitude: asymmetry begins to grow only at
 magnitudes greater then 22.2 which is greater than the limiting magnitude.
%ZZZ verifier ce chiffre trop(?) petit ZZZ
}
 \label{fig:as_mag} 
 \end{figure}

 %\begin{figure}
 %\centering
   %\resizebox{\hsize}{!}{\includegraphics{C_versus_mu.eps}}
 %\caption{Variations in the concentration values for 100 simulated galaxies when changing the limiting surface brightness from $\Delta \mu=1$. The horizontal line shows the mean value of the distribution, which is $\delta C=0.06$.%ZZZ verifier ce chiffre trop(?) petit ZZZ
%}
 %\label{fig:c_mu} 
 %\end{figure}

As said previously, rotational asymmetry is affected by noise even
after correction. This means that fainter objects might appear more
asymmetric and can thus induce a bias in the number of irregular objects
at the faint end of the sample. To estimate this error, we plotted
the asymmetry parameter for a sample of $1000$ simulated galaxies with
magnitudes ranging between $17<K_s<23$ (Fig.~\ref{fig:as_mag}). The plot
shows that asymmetry begins to be affected by noise only at magnitudes
greater than $22.2$, which is the magnitude limit of our working sample. 
In summary, we found the location of the irregular/peculiar objects by simulating a set of regular galaxies  and defining the peculiar area as the area outiside. Then, we plotted the observed data on this plane and count the galaxies in this peculiar area. We counted 10 observed objects in this zone, i.e. $12\%$ of the sample. We can attempt to quantify the error in this classification by considering the regular simulated objects that fall in the irregular zone. This gives the fraction of regular objects that are misclassified. We counted 27 objects out of 1000. we conclude that $12\%\pm2.7$ of our sample corresponds to peculiar 
objects, in the magnitude range $17<K_s<22$.
%Asuming that the error is symmetric, which means that the same fraction of irregular objects is classified as regular, 

\subsection{Regular objects: disk dominated - bulge dominated Separation}
\subsubsection{C-A morphology}
\label{sec:ca}
\begin{figure*} 
\includegraphics[width=0.48\textwidth,height=0.4\textwidth]{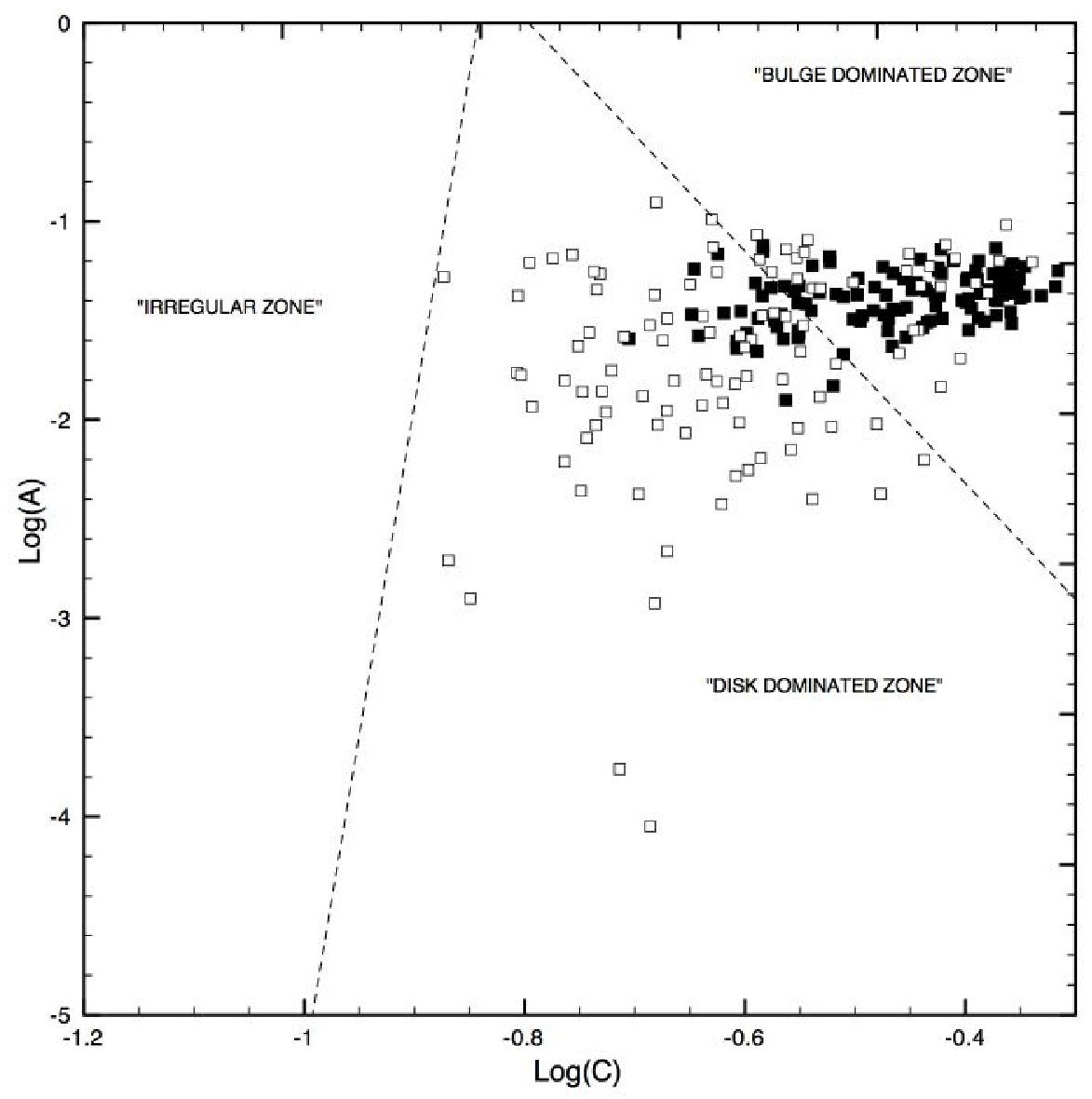}
\includegraphics[width=0.48\textwidth,height=0.4\textwidth]{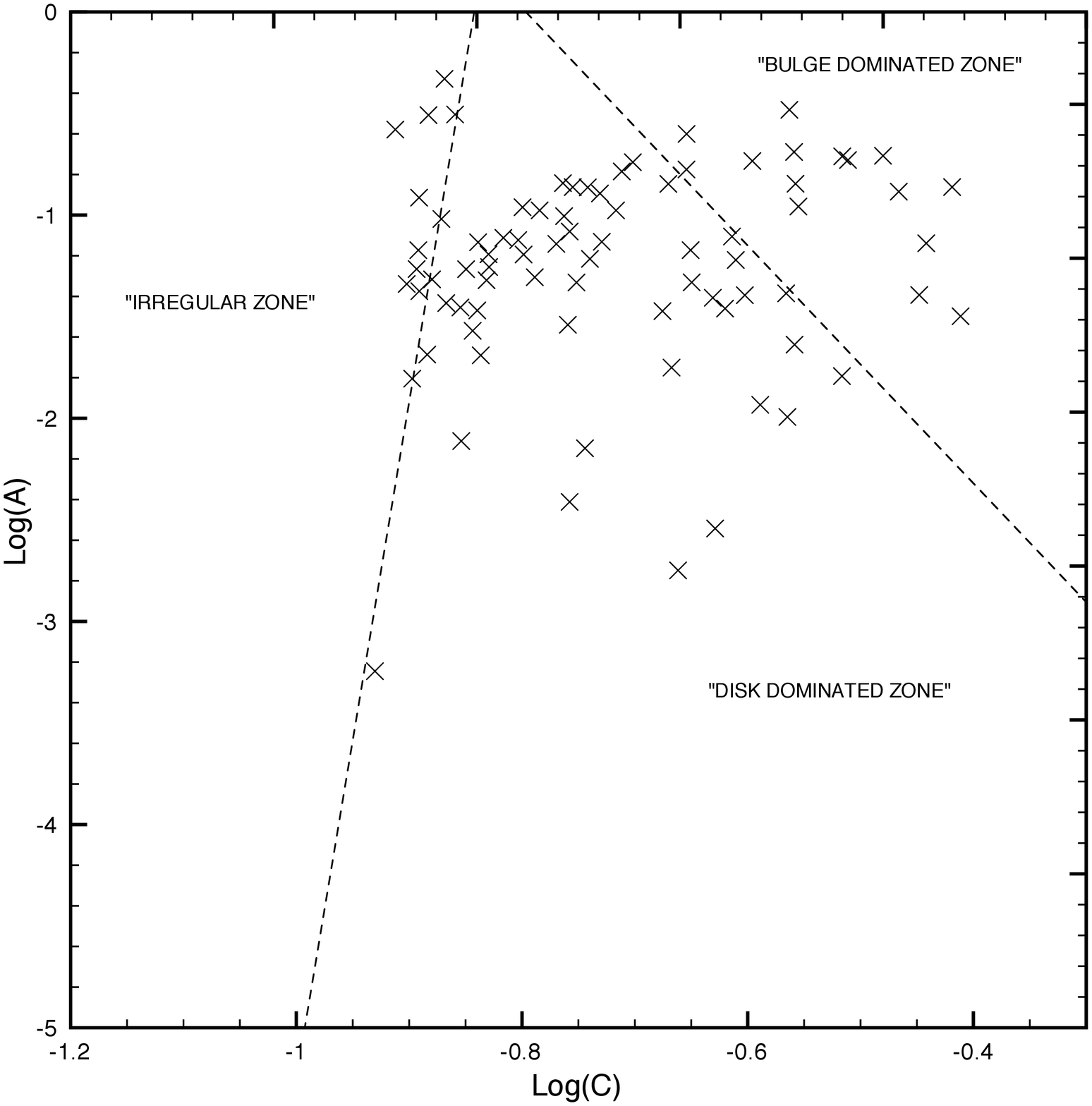}
\caption{C-A calibration and classification. Boundaries are drawn
using an automated classification method (SVM) that avoids the
use of a nearby sample and subjective visual classifications. Left:
simulated objects, open squares: objects
with $B/T<0.2$, filled squares: objects with $B/T>0.8$. Right: real objects. 
%ZZZ Pb: les 2 figures sont identiques. A droite, representer
%uniquement les croix des vrais objets ZZZ
.} \label{fig:CAS} \end{figure*}

The positions of galaxies in the C-A plane are used to separate
the bulge and disk-dominated galaxies as follows:

%The concentration was estimated
%using the iso-ellipses described in \cite{Ab96} at a normalized radius
%of 0.3. Asymmetry was obtained as explained in Section~\ref{sec:irr}. 

 \paragraph{Calibration:}
A calibration of the C-A plane is needed before classification to
investigate where the objects exactly fall. The strategy followed
in this paper is two-fold: first we draw the irregular border as
defined in Section~\ref{sec:irr}. The border between bulge-dominated and disk-dominated objects can be deduced in a more automated way thanks to
the analysis of simulated galaxies. We took the same $1000$ galaxies
as above, for which the morphological type is known, and draw their
positions in the C-A plane (Fig.~\ref{fig:CAS}).  The border is then
defined with a classification method based on support vector machines
\citep{Vap95} \footnote{%The free toolbox from Steve Gunn was used,
http://www.isis.ecs.soton.ac.uk/resources/svminfo/ }. We decided not to use classical boundaries employed in previous works because: a) those boundaries were not obtained in the K-band and b)we were looking for an objective method that did not require visual inspections. Support vector
machines (SVM) non-linearly map their n-dimensional input space into a
"high dimensional feature space". In this high-dimensional feature space,
a linear classifier is constructed. SVM have two main parameters that
can be changed: the kernel function and the tolerance C. The kernel
function corresponds to the expected shape of the border, for instance, 
if the objects are distributed with a gaussian distribution, then a gaussian
kernel will be used. The adjustment of the border will also take a tolerance factor C into account. If C is high, the machine will not allow
any object to be on the wrong side of the border. As a consequence, if
the objects are strongly mixed in a given plane, the border can have a
very complex shape.  In contrast, if C is too low the machine will
not reach an optimal separation.  In a first approach, a linear kernel
was used, assuming that the two families of objects can be separated with
a linear function, in order to be coherent with previous works. Thus the C parameter is set to be infinite. The
results of this separation are displayed in Fig.~\ref{fig:CAS}. 

 \paragraph{Accuracy:}
Automated classifications are useful because they allow a characterization
of errors. Once the boundaries are drawn, we generated another
set of 500 fake objects with known morphology that we place again in the
C-A plane and that we classifed in the three morphological types we have
defined. We then compared the results of our classification scheme with the
initial morphological type. Errors were estimated in magnitude bins,
from $K_s=18$ to $K_s=22$ (Table~\ref{tbl:perfSVM}). We achieved $70\%$
of good identifications up to $K_s=21$ ($AB=23$) and 66\%\ up to
our magnitude limit $K_s=22$. That means that we are able to classify
galaxies in the three main morphological types with a reliable accuracy
at least up to $K_{AB}=23$.  
%With this method we cannot however separate
%intermediate morphological types.

 \begin{table} \centering 
 \begin{tabular}{c c} \hline \hline \noalign{\smallskip} 
Magnitude & Correct Identifications\\ 
\noalign{\smallskip} \hline \noalign{\smallskip}
  $K_s<19$ & $80\%$\\
  $K_s<20$ & $73\%$\\
  $K_s<21$ & $70\%$\\
  $K_s<22$ & $66\%$\\
\noalign{\smallskip}\hline
\end{tabular}
  \caption{Error estimates of the C-A classification. Fraction of misclassified objects for several magnitude ranges. }
  \label{tbl:perfSVM}
\end{table}

 \paragraph{Some words about the C-A plane: }
At first sight, the distribution of galaxies in our C-A plane looks significantly different that what has been reported in previous works using this techniques. \citep{Ab96,Bri98} . Indeed, the slope of the separating border between bulge and disk-dominated galaxies
has been found to be positive, whereas the one found here is negative, although previous classifications are somewhat arbitrary. As a consequence, bulge-dominated galaxies lie in the top right corner of the plane in our classification rather than in the bottom right corner in most other studies.
There might be several reasons to explain this effect:
 \begin{itemize} 
 \item As NACO images are undersampled, highly concentrated
 objects only cover a few pixels, consequently, even a small mismatch
 in the determination of the rotation center can lead to large asymmetries. To check this effect we computed concentration and asymmetry
 parameters for the detected point-like sources. We find  that, indeed, they appear to have higher  asymmetric values than the galaxies.
 \item It might also be a consequence of the method used for the boundary
 estimate. \cite{Ab96} and \cite{Bri98} used a visual inspection
 based on a local survey and, in order to account for the effects of
 redshift, they applied corrections to the concentration parameter.
 In this work, we used a fully automated method based on simulations that
 reproduce the observational conditions very closely and on learning
 classification methods, so no correction is needed in principle,
 but this gives a different border. The question that
 consequently arises is whether the differences in the boundaries produce
 important differences in the classification procedure. All this points
 are fully discussed in Sect.~\ref{sec:comparisons} thanks to
 detailed comparisons with space observations of the sample.
 \end{itemize}

   %\begin{figure}
 %\centering %\includegraphics[scale=0.2]{C_A_plane_stars.eps}
 %\caption{C-A plane for point sources. Empty squares: Disk dominated
 %objects, Filled squares: bulge %dominated objects. Crosses: point like
 %sources.} %\label{fig:asym_stars} %\end{figure}

\subsubsection{Model-fitting morphology}
\label{sec:gim2d}
The second method is based on a direct two
components fitting with exponential and de Vaucouleurs profiles, using
GIM2D \citep{Sim02}. The 2D galaxy model used by GIM2D has 11 parameters
that are fitted to the real data. The most important ones are the total
flux and the bulge fraction $B/T$ ($=0$ for pure disk systems). Other
parameters are the (i)disk scale radius $r_d$, (ii)the disk inclination $i$,
(iii)the effective radius $r_e$ (iv)the ellipticiy $e$ of the bulge component,
and other geometric parameters for the center and orientation of
both components. As GIM2D also estimates the local sky level using image statistics, this is the value that is used.

\paragraph{PSF reconstruction:}
To obtain reliable results, GIM2D needs a noise-free,
well-sampled PSF. This is why special attention has been paid to PSF
reconstruction. Here, classical methods, such as DAOPHOT or Tiny Tim,
could not be used for two reasons. First, the Adaptive
Optics PSF has a specific shape that is neither 'seeing limited like'
nor 'spatial like'.  An AO system operated with a guide star of moderate
brightness ($V\sim 14$ ) can only partially correct for turbulence-induced
distortions. This partially-corrected PSF consists of two components:
a diffraction-limited core, superimposed on a seeing halo. Second, to have a larger field and a better sensitivity, data were
under-sampled by a factor 2 (0.054\arcsec\ pixel scale, whereas
0.02\arcsec\ is needed
to be Nyquist sampled).

 \begin{figure}
 \centering
   \resizebox{\hsize}{!}{\includegraphics{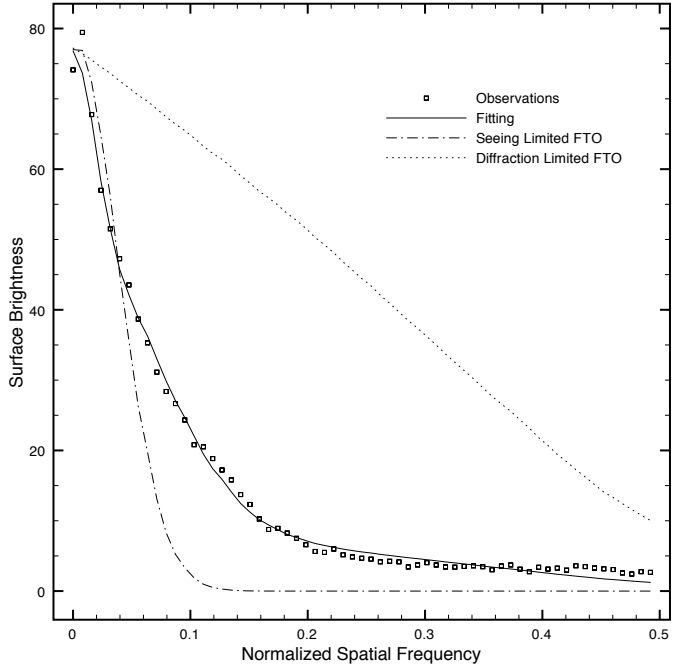}}
 \caption{Example of PSF fitting in Fourier space. Squares:
 observations, dashed line: seeing-limited MTF, dotted line: diffraction-limited MTF. The AO MTF contains higher frequencies than the seeing-limited one. The telescope diffraction limit is not reached
however in this example due to the undersampling of the instrumental setup.}
 \label{fig:psf_fit} 
 \end{figure}

We developed a simple algorithm that uses field stars to generate
Nyquist-Shannon-sampled PSFs by means of a fitting procedure in Fourier
space. The process is as follows: we generate an artificial PSF with
a diffraction-limited core and a Gaussian halo, with the distribution
\[
\textrm{PSF}_{art} (x,y) = k \ \left[\textrm{SR} \times A (x,y) 
+(1-\textrm{SR}) \times \exp \left(-\frac{x^2+y^2}{\sigma^2}
\right)\right]
\label{eq:PSF}
\]
where SR is the Strehl ratio, $A(x,y)$ the bi-dimensional Airy
function, $\sigma$ the Gaussian dispersion that can be related to the
seeing and $k$ is a normalization factor.  This artificial PSF is built
with a Nyquist-Shannon sampling,  binned by a factor 2 to reach the real-image
pixel scale and finally Fourier-transformed to create a simulated MTF
(power spectrum). On the other hand, for each observed star,
its Fourier transform is fitted with the simulated MTF. The parameters
estimated that way (SR, $\sigma$, and $k$) are then used to build an
estimate of the PSF with the correct Nyquist-Shannon sampling.
 % \begin{figure} %\centering 
 %\includegraphics[scale=0.8]{fit_algo.eps}
 %\caption{Fitting algorithm} 
 %\label{fig:psf_fit_algo} 
 %\end{figure}

Working in Fourier space avoids including the background estimate and
PSF position as a fit parameter, which is particularly delicate in our
case, since the FWHM is less than 2 pixels large. 
%As a consequence,
%the fitting is done with only 3 parameters: an estimate of the
%Strehl ratio ($SR$), that measures the fraction of coherent flux,
%the gaussian dispersion ($\sigma$) that can be seen as an estimate of
%the seeing and a normalisation factor ($k$). 
In the few cases where the fitting procedure does not converge a second
Gaussian halo is added.  Fig.~\ref{fig:psf_fit} shows the result of
the fitting for one star in the spatial frequency domain. In this paper,
we do not consider variations in the PSF caused by adaptive optics, such
as anisoplanetism, but we are working on building a complete model for
PSF estimate for future observations.

\paragraph{Error analysis:}
%\begin{figure*}
%\centering
%\includegraphics[width=0.48\textwidth]{STAR1.eps}
%\includegraphics[width=0.48\textwidth]{STAR1_res.eps}
%\caption{Original (left panel) and residual (right panel). We have subtracted the GIM2D models to
%the original image and look for residual objects.  Left: original image,
%Right: image after model substraction. Only the objects
%identified as point sources remain.}
%\label{fig:res} 
%\end{figure*}

\begin{table}
\begin{tabular}{c c c} 
\hline \hline \noalign{\smallskip} 
%\multicolumn{2}{c}{NACO} & & \multicolumn{2}{c}{ACS} \\
Original & Model & Residual\\  
\noalign{\smallskip} \hline \noalign{\smallskip}
\includegraphics[scale=2]{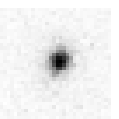}  & \includegraphics[scale=2]{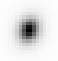} & \includegraphics[scale=2]{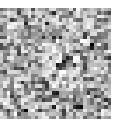} \\
\hline \noalign{\smallskip} \includegraphics[scale=2]{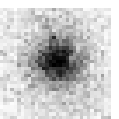} & \includegraphics[scale=2]{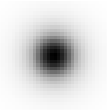} & \includegraphics[scale=2]{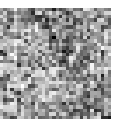}\\
\hline \noalign{\smallskip}
\includegraphics[scale=2]{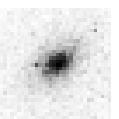} & \includegraphics[scale=2]{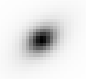} & \includegraphics[scale=2]{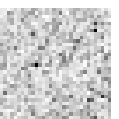} \\
\hline \noalign{\smallskip}
\includegraphics[scale=2]{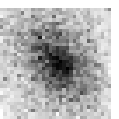} & \includegraphics[scale=2]{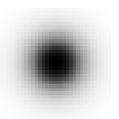} & \includegraphics[scale=2]{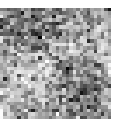} \\
\includegraphics[scale=2]{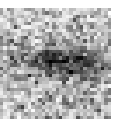} & \includegraphics[scale=2]{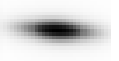} & \includegraphics[scale=2]{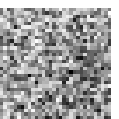} \\
\includegraphics[scale=2]{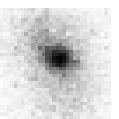} & \includegraphics[scale=2]{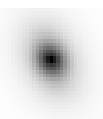} & \includegraphics[scale=2]{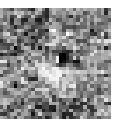} \\
\noalign{\smallskip}\hline
\end{tabular}
\caption{GIM2D output for some objects. Left: original images, Middle: GIM2D models, Right: Residuals. The small thumbnails show
the real and the model galaxy. The image size is $1.7^{'}\times1.7^{'}$} \label{tbl:GIM2_output} \end{table}

We ran GIM2D on the 79 objects with magnitudes ranging between
$K_s=17-22$, using a two components model and the artificial
PSF built as described above. We used the GIM2D mode that allows use of oversampled PSFs to deal with undersampled data, since the PSF recovered with the method explained above is Nyquist sampled.The fitting converged for the whole
sample, and the results are quite convincing in terms of residual
images. (Table ~\ref{tbl:GIM2_output})
%Fig.~\ref{fig:res} shows the result of substracting the models
%to the original image. It can be seen that only the three objects
%identified as stars in Fig.~\ref{fig:field} remain in the residual
%image. 

%We therefore consider that reliable morphological information
%can be obtained up to the magnitude limit $K_s=22$. 

%ZZZ est comparable
%a SWAN? ZZZ

%\begin{figure*}
 %\centering
% \includegraphics[width=12cm]{GIM2D_models.eps} 
 %\caption{GIM2D models for one field. The small thumbnails show
 %the real and the model galaxy. For bright galaxies there is a good
 %agreement. For the faintest ones it is difficult to estimate the accuracy
 %of the model. This is why systematic error estimations are necessary.}
 %\label{fig:gim_mod} \end{figure*}

Visual inspection of the models compared to the real images
(Table~\ref{tbl:GIM2_output}) also reveals good agreement, in particular for
bright sources. For the faintest objects, however, it is more difficult
to estimate the fitting accuracy. Indeed, inspection of image residuals is not a robust accuracy test, since there may still be strong degeneracies even when the image residuals do not show any features.This is why objective and systematic
error characterization is needed. To do so we generated a sample of $1000$
synthetic galaxies with known galaxy parameters uniformly distributed:
$0<B/T<1$, $0<r_d<0.5\arcsec$, $0<r_e<0.5\arcsec$, 
$0<e<0.7$, $0\degr<i<85\degr$. The Sersic bulge index was fixed at $n=4$, and both bulge and disk
position angles were fixed to $90\degr$. As explained in \cite{Sim02},
the goal of these simulations is to characterize biases and error.
The uniformity of the parameter distributions adopted here is therefore
perfectly suitable even though real galaxy parameters may not be so
distributed. Each simulation is convolved with the reconstructed PSF. The
same PSF is used in both creating and analyzing the simulations, so
the results will not include any error due to PSF mismatch. In order to
simulate background noise, objects are embedded at random positions in the
fields and detected with the same Sextractor parameters as for the real
sources. Finally, the GIM2D output files are processed through the same
scripts to produce a catalog of final recovered structural parameters.

\begin{table*}
\begin{center}
\begin{tabular}{c|ccc|ccc|ccc|ccc|}
\hline\hline\noalign{\smallskip}
% & \multicolumn{12}{c|}{$log(r_{hl}\arcsec)$} \\
 & \multicolumn{3}{c|}{$-1 < \log (r_{hl}) < -0.75$} & 
            \multicolumn{3}{c|}{$-0.75 < \log (r_{hl}) < -0.5$} &
            \multicolumn{3}{c|}{$-0.5 < \log (r_{hl}) < -0.25$} & 
            \multicolumn{3}{c|}{$-0.25 < \log (r_{hl}) < 0$}  \\
Magnitude & $\overline{\Delta B/T}$ & $\sigma \Delta B/T$ & N & $\overline{\Delta B/T}$ & $\sigma \Delta B/T$ & N & $\overline{\Delta B/T}$ & $\sigma \Delta B/T$ & N & $\overline{\Delta B/T}$ &
$\sigma \Delta B/T$ & N \\
\noalign{\smallskip}\hline\noalign{\smallskip}
$[17-17.5]$ & --0.110 & 0.247 & (19) & --0.034 & 0.140 & (22) & --0.002 & 0.056 &
(4) & --0.030 & 0.060 & (8) \\
$[17.5-18]$ & 0.165 & 0.088 & (4) &  0.037 & 0.277 & (18) & 0.027 & 0.157 &
(23) & 0.010 & 0.110 & (11)  \\
$[18-18.5]$ &  --0.002 & 0.141 & (19) & 0.020 & 0.262 & (58) &  0.162 & 0.174 &
(72) &  0.226 & 0.155 & (28)   \\
$[18.5-19]$ & --0.114 & 0.324 & (14) & 0.021 & 0.309 & (71) &  0.213 & 0.196
& (83) &  0.170 & 0.159 & (30)   \\
$[19-19.5]$ & --0.064 & 0.221 & (20) & 0.212 & 0.266 & (90) &  0.192 & 0.252 &
(51) & 0.145 & 0.027 & (3)   \\
$[19.5-20]$ & 0.100 & 0.259 & (24) & 0.270 & 0.298 & (110) &  0.181 & 0.258
& (34) &  0.140 & 0.100 & (6)   \\
$[20-20.5]$ & 0.105 & 0.430 & (36) & 0.163 & 0.332 & (107) &  0.113 & 0.281 &
(13) &  N/A & N/A & (0)   \\
$[20.5-21]$ & 0.050 & 0.476 & (56) & 0.148 & 0.351 & (81) &  0.194 & 0.054 &
(3) &  N/A &  N/A & (0)   \\
\noalign{\smallskip}\hline
\end{tabular}
\end{center}
\caption{Error analysis of the bulge fraction B/T for different recovered
magnitude ranges and different bins in recovered galaxy size. The galaxy size
is represented by the half-light radius and is distributed into 4 bins
in $\log (r_{hl})$. In the top left corner bright and
small objects are found whereas faint and large objects are placed in the
bottom right corner. $\overline{\Delta B/T}$ is the average difference between introduced and
recovered values of B/T, while $\sigma \Delta B/T$ is the dispersion (see text for details). 
N is the number of simulations used for each bin. }
\label{tbl:BT}
\end{table*}

Following the \cite{Sim02} procedure, we decided to represent errors
in a set of two-dimensional maps giving systematic and random errors at
each position. The GIM2D parameter space is a complex space with 11
dimensions, so these maps can only offer a limited representation of the
complex multidimensional error functions but makes interpretation much
simpler. The error analysis presented in this paper focuses
on the error made on the main morphological estimator, the bulge fraction,
as a function of two main parameters: apparent magnitude and half light
radius. Systematic errors are computed as the mean difference between the introduced and the recovered value: $\overline{\Delta B/T}=\frac{\sum({B/T}_i-{B/T}_r)}{N}$ and random errors as the square root of the variance of the difference: $\sigma \Delta B/T=\sqrt{\frac{\sum{(\Delta B/T-\overline{\Delta B/T})^2}}{N-1}}$. Table~\ref{tbl:BT} precisely shows in details the sources of
errors on B/T as a function of galaxy magnitude and half-light radius.

The main result after looking at the results of simulations is that, for objects brighter than $K_s\sim19$ ($AB\sim 21$), the bulge fraction is recovered with a bias close to zero and a random error around $20\%$. This is true even for small objects ($-1<log(r_{hl})<-0.75$), and it is comparable to what is obtained for the brightest objects in the I-band with HST \citep{Sim02}. For fainter
magnitudes, we can see two main effects:

\begin{itemize}

\item first, an increasing bias in the bulge fraction ($\sim 20\%$), in particular for large objects ($log(r_{hl})> -0.75$). 
%\textbf{random} errors rise significantly \textbf{($\sim30\%$)} 
%which means that we
%underestimate bulge fractions. 
This is a well-known GIM2D effect for
low S/N objects \citep{Sim02}: the outer wings of steep surface
brightness profiles, such as the $r^{1/4}$ profile, are hidden in low S/N objects which artificially decreases the recovered
bulge fraction. This can also be a consequence of errors in the sky-level estimate that causes that a fraction of the flux is hided by noise.

\item second, a slight increase in the random error, which becomes closer to $30\%$. This is a consequence of an increasing degeneracy of the parameters space with decreasing S/N.

\end{itemize}

 But for most of the
objects brighter than $K_s=19$ galaxy parameters can be
estimated correctly ($\sigma \sim 0.2$ and b $ < 0.1$), even for 
small objects ($r_{hl} < 0.3\arcsec $) of size comparable the limits
of space observations (see Section~\ref{sec:comparisons}).

\subsection{Results of the analysis and comparison of classifications}
We classified the galaxies into three main morphological types according
to the fitting results. One of the main results is that about $12\%\pm2.7\%$
of our sample corresponds to peculiar or irregular 
objects (10 objects out of 79).
For the rest of the sample, the GIM2D analysis finds 21 ($\sim26\%$) bulge-dominated galaxies ($B/T > 0.5$) and 48 ($\sim60\%$) disk-dominated ($B/T < 0.5$) while for the C-A classification, we find
54 ($\sim67\%$) disk-dominated galaxies and 15 ($\sim19\%$) bulge-dominated ones.

% and a slighly
%higher fraction of irregular galaxies ($\sim 30\%$). 

\begin{figure}
 \centering
  \resizebox{\hsize}{!}{\includegraphics{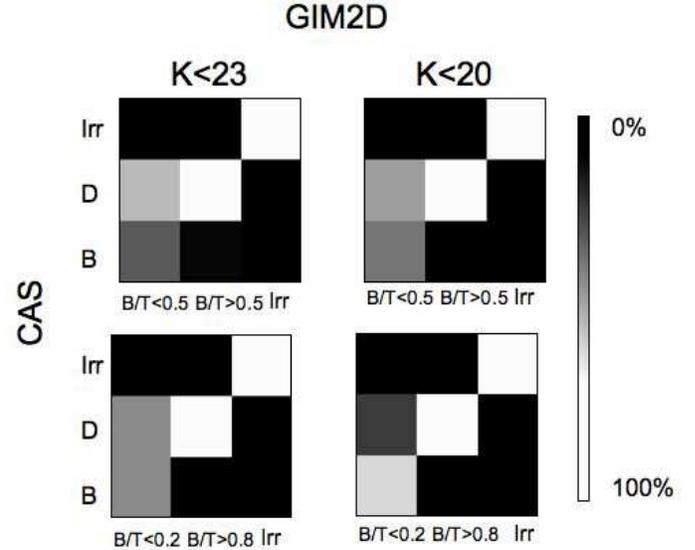}}
 \caption{Comparison of classification methods, show the
 probability that a galaxy classified with GIM2D is classified in the same
 morphological type by C-A. (see text for details).} \label{fig:m_comp}
 \end{figure}

Looking in more detail into the relaibility of the two
classification schemes, we did a one-to-one comparison of the
morphological types assigned by the GIM2D analysis or the C-A one
(Fig~\ref{fig:m_comp}): we computed 
the probability that a galaxy classified using the GIM2D classification
is classified with the same morphological type by C-A. The probability was computed by dividing the number of galaxies in each morphological C-A
bin by the total number of galaxies of the same type selected with GIM2D.
Overall, there is good agreement between both classifications in
the whole sample. The probability that a disk dominated galaxy identified by
GIM2D has the same morphological type in C-A classification is $p=0.81$,
but only $p=0.30$ for bulge-dominated galaxies including the faintest objects
($K_s<23$). For irregulars, it is obviously $p=1$ since the detection
procedure is the same in both methods. 

%ZZZ Alors pourquoi il y a
%30\%\ d'irregulieres avec CAS et seulement 25\%\ avec GIM2D??? a
%clarifier ZZZ

There might be two reasons why the classifications are not exactly
the same. First, the S/N might cause discrepancies. Indeed, as we
show in Sect.~\ref{sec:gim2d}, at low S/N, GIM2D tends to under estimate
the bulge fraction. This implies that some galaxies detected by GIM2D as disk-dominated are in fact detected as bulge-dominated by C-A. Figure~\ref{fig:m_comp}
shows the effect of reducing the limiting magnitude to $K_s=20$: the
fraction of objects classified as bulge dominated by GIM2D and C-A rises
up to 0.67. Second, it might be  a problem of definition. Indeed, the
morphological bins are not exactly the same in both classifications. In
particular, objects with intermediate morphological type (i.e $B/T\sim$0.5)
might cause discrepancies. If we remove those objects from the sample, $80\%$ of the bulge-dominated objects and $95\%$ of the disk-dominated objects detected
by GIM2D are also detected by C-A with the same classification.

Either way, the comparison of both classifications allows a quantification of the error in classification of regular galaxies in the sense that it seems reasonable to think that the true value should be somewhere between the two results. The GIM2D estimate thus gives a lower limit for the early-type fraction and C-A the upper-limit and vice versa for the late-type fraction.  This way, we conclude that the mixing of population in our sample is:  $24\%\pm4\%$ of early-type galaxies, $64\%\pm4\%$ of late-type galaxies,  and $12\%\pm2.7\%$ of irregular/peculiar galaxies.

%   \begin{figure}
% \centering \includegraphics[scale=0.3]{morpho_histos.eps}
% \caption{Comparison of classifications. Distribution of morphological
% types in our sample as obtained as with the two different methods
% explained in the text. There's an overall good agreement between both
% methods. Error bars show poissonnian errors.} \label{fig:morpho_histo}
% \end{figure}

\begin{figure}     
\centering          
\begin{tabular}{c c c}     % 7 columns 
      
                      % To combine 4 columns into a single one 
 %\multicolumn{2}{c}{Method\#3}\\ 
\hline disk dominated & bulge dominated & Irregular \\
\hline \includegraphics[scale=1.5]{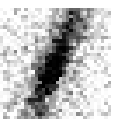} &
\includegraphics[scale=1.5]{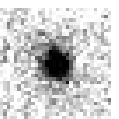} &
\includegraphics[scale=1.5]{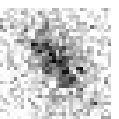} \\
%  \hline 
$z=0.22$ & $z=0.76$ & $z=0.04$ \\
  \hline \includegraphics[scale=1.5]{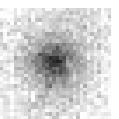} &
\includegraphics[scale=1.5]{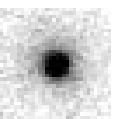} &
\includegraphics[scale=1.5]{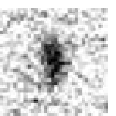} \\
% \hline 
$z=0.96$ & $z=0.98$ & $z=1.28$ \\
 \hline \includegraphics[scale=1.5]{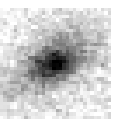} &
\includegraphics[scale=1.5]{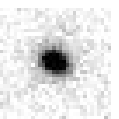} &
\includegraphics[scale=1.5]{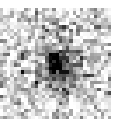} \\
% \hline 
$z=1.09$ & $z=1.37$ & $z=0.77$ \\
\hline      
\end{tabular}
\caption{Example of classification in the three main morphological types at different redshifts. The image size is $1.7^{'}\times1.7^{'}$}
\label{fig:class_ex}
\end{figure}

Our results offer the first direct measurement of the distribution of galaxy in three morphological types at $z\simeq1$ from high spatial-resolution imaging in the K-band. We observe that the fraction of $12\%$ of irregular objects at $z=1$ is significantly higher than the fraction of these objects in the local Universe, confirming from rest-frame data at $\sim1$ microns the well documented trend of this population increasing with redshift (e.g. \cite{Bri98}). However, this result must be taken with 
caution. Indeed GIM2D accuracy decreases for objects fainter than $K_s=19$, which represents $80\%$ of the
sample. Moreover at the faint end, the fraction of irregular objects can
be overestimated because of the low S/N. 
%In addition there can be a mismatch between local and high
%redshift morphological classifications due to the ``morphological
%k-correction'' (see Section~\ref{sec:hst}). 
But there are good reasons to
consider this result significant. Even though there is an over estimation
of disks in the faint end, the morphological classification bins are large
enough to reduce the number of false classifications. Indeed, even in
the zones where the random error in the bulge fraction estimate is
$\sim 0.3$ or larger, we do not classify a pure bulge ($B/T\sim1$) as a disk.

%\section{Morphological Evolution}

%We divided the 60 matched objects sample into 2 categories: low redshift ($z_{phot}<0.8$) and high redshift objects ($z>0.6$) and plot the evolution for each morphological type and for both classification methods (CAS and GIM2D). Figure~\ref{fig:morpho_evol} shows the counts of each morphological type per square degree and per redshift bin.  Error bars show poisson errors. \\

%VERY STRANGE RESULTS:

%\begin{itemize}

%\item The fraction of irregulars seems to rise

%\item However: we find less spirals and more ellipticals at high redshifts than at low redshifts. selection effects? classification mismatch? Too few objects? Don't really know...

%\end{itemize} 

%\begin{figure*}     
%\centering          
%\begin{tabular}{c c}     % 7 columns 
      
                      % To combine 4 columns into a single one 
 %\multicolumn{2}{c}{Method\#3}\\ 
%\hline \includegraphics[scale=.3]{morpho_evol_CAS_NC.eps} & \includegraphics[scale=.3]{morpho_evol_GIM2D_NC.eps}\\  
             
%\end{tabular}
%\caption{Morphological Evolution. left: CAS, right: GIM2D}
%\label{fig:morpho_evol}
%\end{figure*}

\section{Comparison with ground-based and HST observations} % and evolution
%considerations}
\label{sec:comparisons}
In this section we compare our AO observations with ground-based and space
observations.

\subsection{Ground-based observations}
Effective radii of local galaxies, except for compact dwarf
galaxies, range from $\sim 1.0$ to $\sim10$ kpc depending on their
luminosity \citep{Ben92,Imp96}.  Our spatial resolution of
$\sim0.1\arcsec$
corresponds to about 1 kpc at $z\sim1$ and we should be able to
determine morphological types even at $z>1$.
Thus, in order to estimate the performance of AO deep imaging and to
justify the automated morphology classification, we compared our images
with deep I-band seeing-limited images taken from the  
Canada-France-Hawaii Telescope Legacy Survey
(CFTHLS)\footnote{\url{http://www.cfht.hawaii.edu/Science/CFHLS/}}.
One of the so-called deep fields is 
centered on the COSMOS field, although it is smaller than the total
COSMOS area (1 square degree out of 2). Here we used the release
T0003 images (March 2006)\footnote{\url{http://terapix.iap.fr}}, 
especially the deep $i'$ one,
corresponding to a total integration time of 20 hours, with an
average FWHM of $\sim 0.7\arcsec$.

%We first have performed simulations: We artificially redshifted to
%$z\sim1$a pure bulge galaxy with effective radius $r_e\sim1.4kpc$
%($\sim 0.2"$ at $z\sim1$) and a pure disk galaxy with $r_d\sim1.4kpc$
%and convolved them with Megacam and NACO PSFs respectively. Clearly
%the gain in spatial resolution allows a better separation between
%both morphological profiles.
%While it is
%difficult to separate both profiles in the seeing-limited observations,
%the AO ones, reveal a clear separation.

We compared real data by selecting a galaxy classified as a disk
by GIM2D and C-A in the NACO data and by comparing it to the results
obtained with CFHTLS data. %and NACO images respectively and performed
%a fitting to the surface brightness profile using the iso-polar space
%presented in section 4, with a PSF-convolved de Vaucouleurs profile
%and an exponential profile. The figure shows the results for the 2
%galaxies. Again, in the seeing-limited observations it's dificult to
%establish which one is the bettter fit, whereas the NACO data clearly
%reveal that it's likely an exponential profile.
We computed the surface brightness profile within the isophotal area enclosed by pixels $3\sigma$ above the sky level. The corresponding limiting surface brightness is $\mu=20$ $mag.arcsec^{-2}$ for the NACO image and $\mu=25$ $mag.arcsec^{-2}$ for the MegaCam image. Sky levels and the corresponding isophotal areas were both determined using \textsc{SExtractor}.

\begin{figure*}     
\centering          
\includegraphics[width=0.48\textwidth,height=0.4\textwidth]{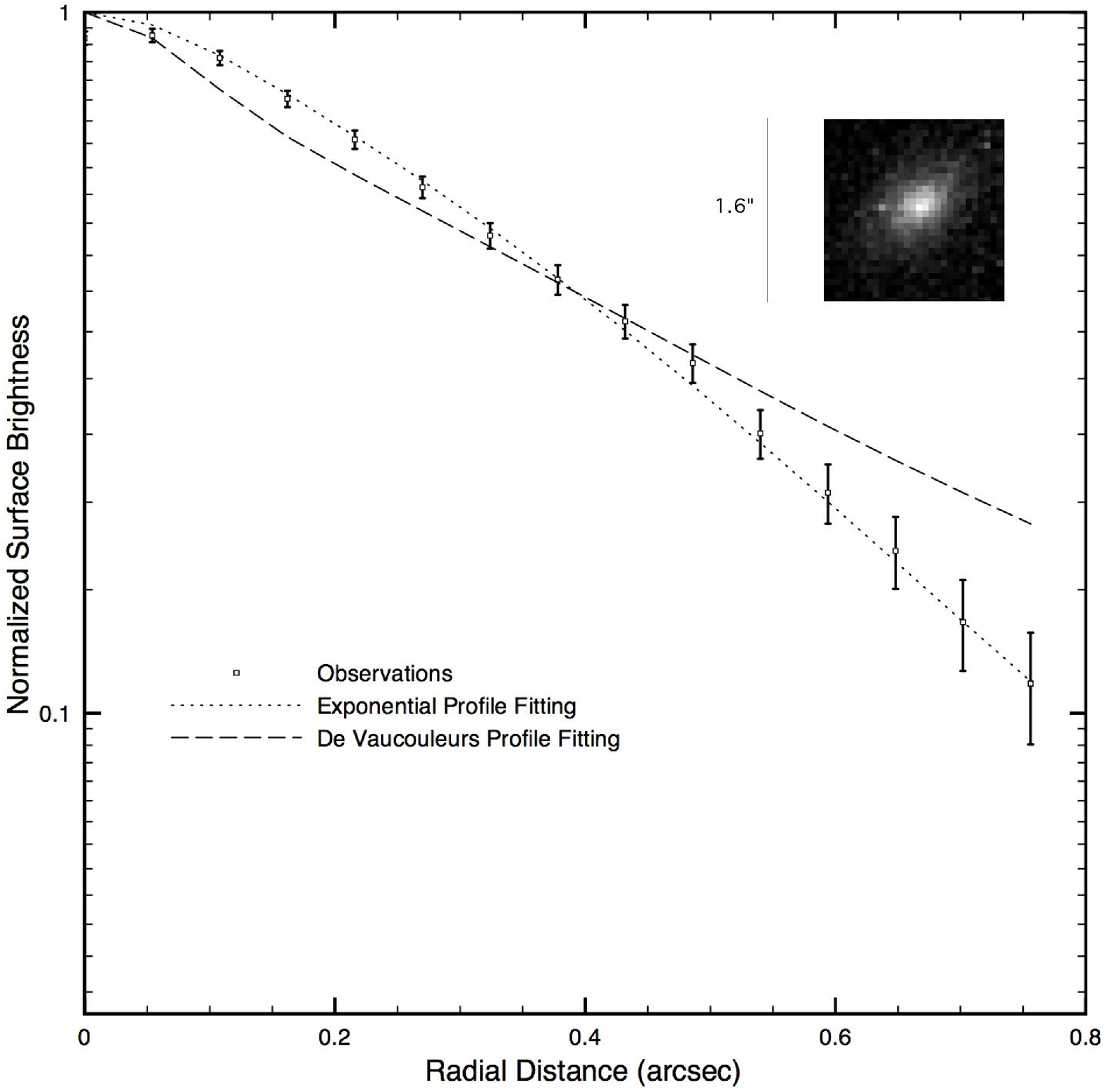} 
\includegraphics[width=0.48\textwidth,height=0.4\textwidth]{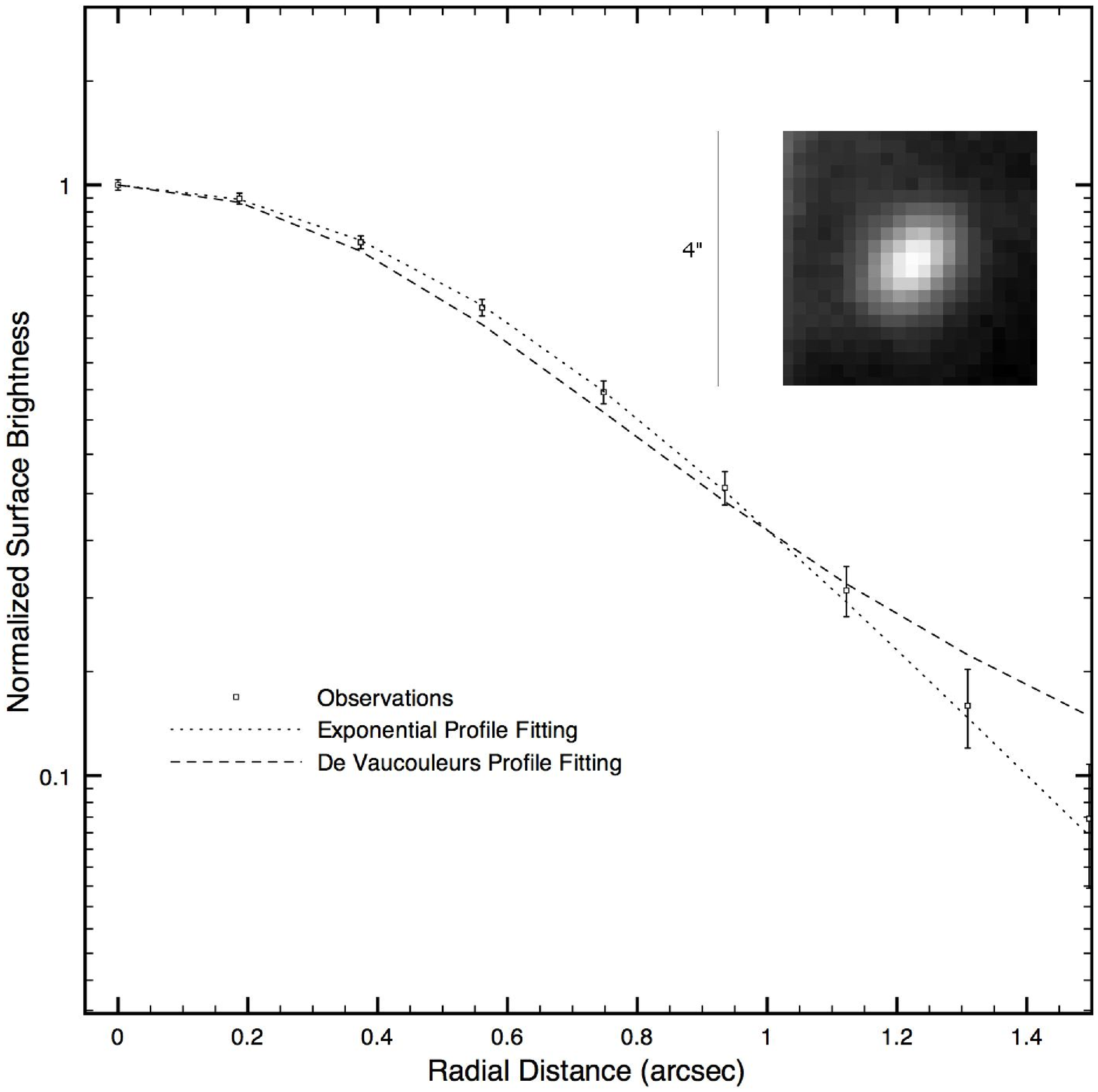}
\caption{Comparison with ground-based observations. We performed a profile
fitting on the same real
galaxy observed with NACO (left) and MegaCam (CFHTLS-$i'$ band, right). The galaxy magnitude is $K_{AB}=20.5$ and $i'=21.3$. Surface brightness profiles were computed within the isophotal areas enclosed by pixels $3\sigma$ above the sky level. The corresponding limiting surface brightness is $\mu=20$ $mag.arcsec^{-2}$ for the NACO image and $\mu=25$ $mag.arcsec^{-2}$ for the MegaCam image. The fit was done with a pure de Vaucouleurs  and exponential
profile. %ZZZ rajouter les
%magnitudes K et I de la galaxie, et une sous-image dans chaque plot,
%de 1.6'' pour NACO et 4'' pour MegaCam (pour montrer la difference
%d'extension dans les 2 filtres).
} \label{fig:prof_fit}
\end{figure*}

The surface brightness profile was fitted with both a 
PSF-convolved de Vaucouleurs profile
and a PSF-convolved exponential profile. Figure~\ref{fig:prof_fit} shows that, with
seeing-limited observations, it is more difficult to establish a
clear separation between both profiles at small distances from the galaxy center (i.e. $\sim 0.2^{"}$), even if the determination 
of the brightness profile is
possible at much larger distance (i.e. $\sim 1^{"}$) thanks to the depth of the images
and the low noise level of the sky background. This supports the results obtained with GIM2D, which show that a correct estimate of the bulge fraction is possible for small objects. Although ultra-deep
sub-arcsecond imaging is powerful in terms of high number statistics,
thanks to the wide field coverage, we consider that it is more rewarding to look at a smaller sample of
galaxies, but with more reliable morphology determinations thanks to the spatial
gain of the AO.

%\begin{figure*}     
%\centering          
%\begin{tabular}{c c}  
%%\begin{figure}   % 7 columns 
    % 
                      %% To combine 4 columns into a single one 
 %%\multicolumn{2}{c}{Method\#3}\\ 
%\hline \includegraphics[width=9cm]{NACO_radial.eps} & \includegraphics[width=9cm]{CFH_radial.eps}\\  
%
   %%\end{figure}             
%\end{tabular}
%\caption{Comparison with ground based observations. We simulated a pure
%bulge galaxy (de Vaucouleurs profile) and a pure disk (Exponential
%profile) noise free as it would be observed by NACO and Megacam
%respectively with the same characteristic radii. (Left: NACO simulated
%data, Right: Megacam simulated data). The two mean radial profiles can be
%easily separated when observed with NACO whereas it's much more difficult
%with Megacam. } \label{fig:prof_comp} \end{figure*}

\subsection{Space observations}
\label{sec:hst}

\begin{figure*}
\includegraphics[width=0.48\textwidth,height=0.4\textwidth]{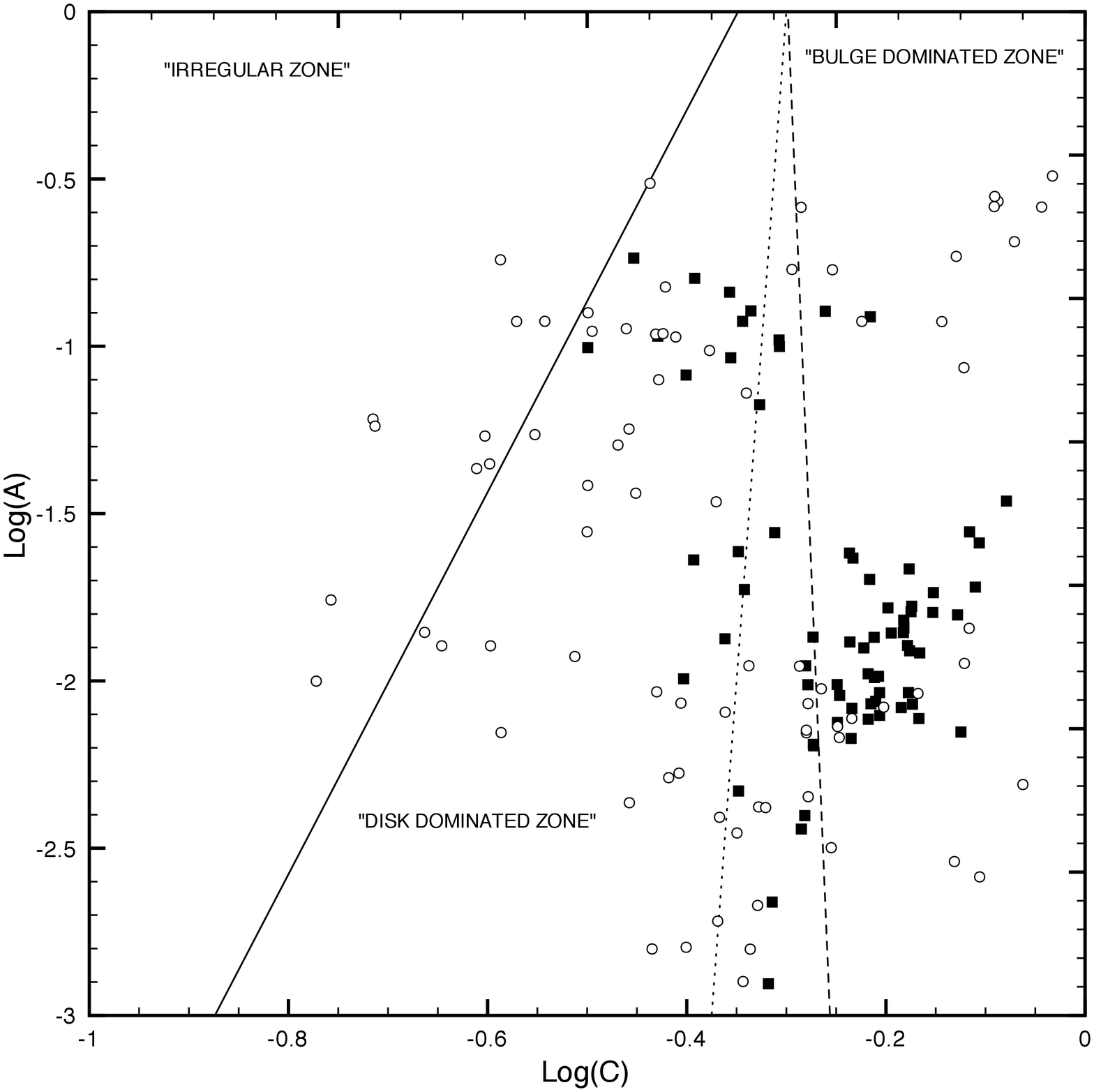} 
\includegraphics[width=0.48\textwidth,height=0.4\textwidth]{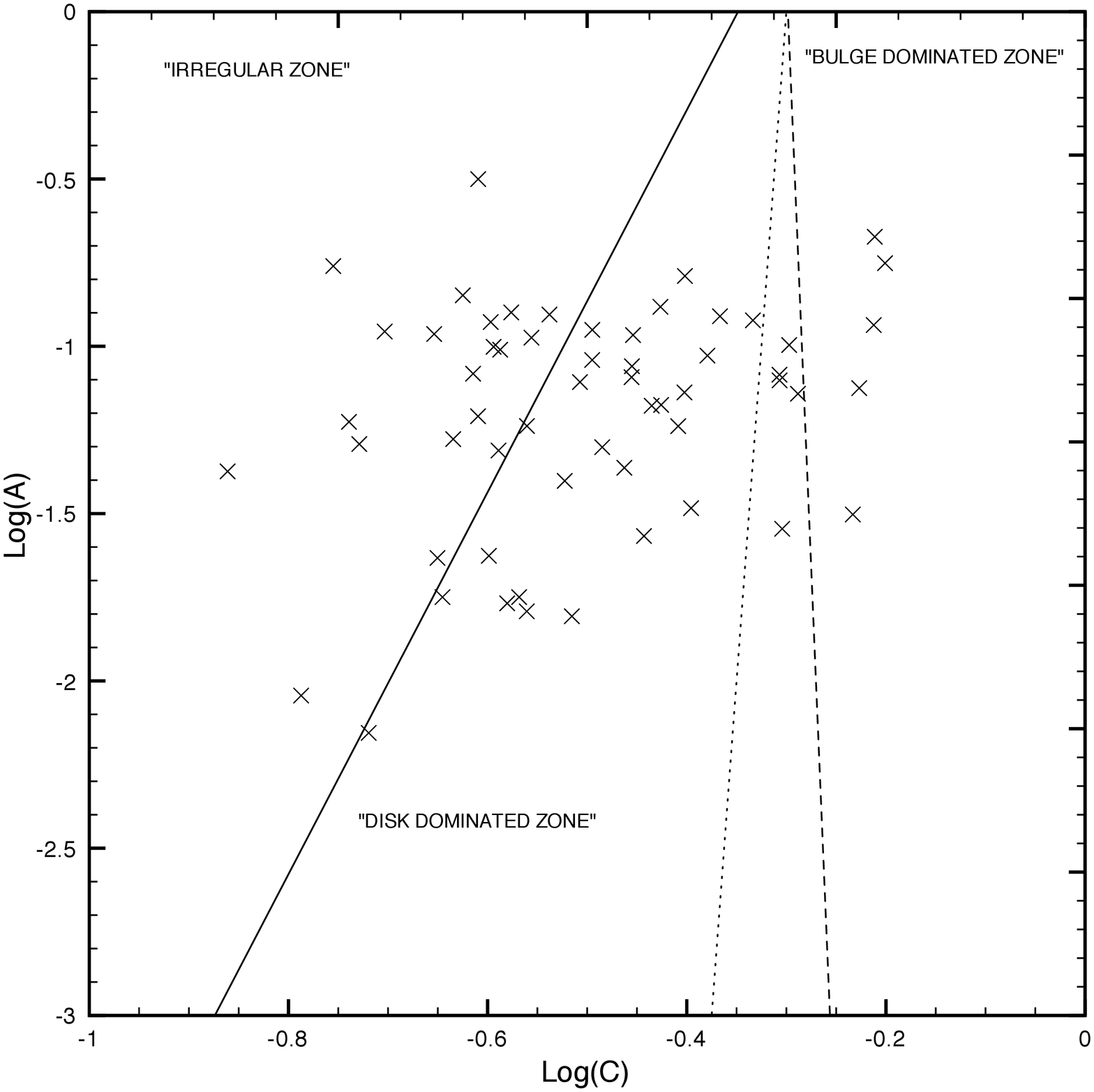}
\caption{C-A cut for the ACS images. The same classification procedure has
been repeated for the same sample observed with ACS in the I-band. Left:
simulated objects. Right: real objects. Circles: $B/T<0.2$,
Filled squares: $B/T>0.8$, Crosses: real objects. Dotted line is the border used in \cite{Bri98} to separate bulge dominated from disk dominated, dashed line is the one computed in the paper.} \label{fig:CAS_ACS}
\end{figure*}

We compare our images with space data taken from the COSMOS survey.  Since our observed fields were selected within the COSMOS area, the same objects were observed with the HST-ACS in the I-band at high spatial resolution. We thus morphologically classified the 60 objects for which the photometric
redshift are known (Sect.~\ref{sec:zphot}). We used those results to both estimate the effect of the observation band on morphology and to validate our method to divide the C/A plane. The C-A
estimators were calibrated with simulations using the same method as for the K
band data. Standard boundaries, from other existing works, were used to divide the C-A plane. Figure~\ref{fig:CAS_ACS} shows the C-A plane cut. The figure also shows the border between bulge-dominated and disk-dominated galaxies obtained with the automatic method described in Sect.~\ref{sec:ca} for this population. We again find a
negative slope for the border between disk and bulge dominated objects. We
find for the whole sample $32\%\pm1.6\%$ irregulars, $47\%\pm1.5\%$ disk-dominated, and $21\%\pm2.5\%$ bulge dominated.

  \begin{figure}
 \centering
   \resizebox{\hsize}{!}{\includegraphics{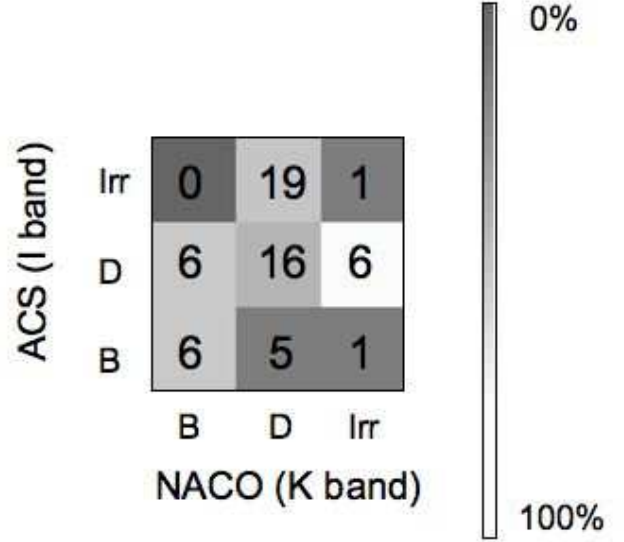}}
 \caption{Galaxy distribution: comparison between K-band and I
 band C-A classifications. The figure shows the probability that a galaxy in the K-band is classified in the same morphological type in the I-band.}
 \label{fig:morpho_histo_ACS_NACO} \end{figure}
 
 \subsubsection{About boundaries}
%ZZZ A revoir ZZZ

\begin{figure}
 \centering
  \resizebox{\hsize}{!}{\includegraphics{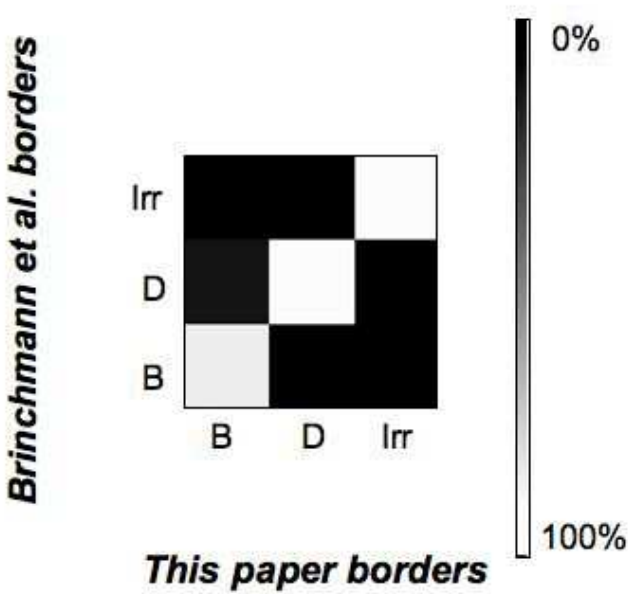}}
 \caption{Comparison of classifications with different boundaries. We
repeat the morphological classification with the boundaries used by
\cite{Bri98}. We conclude that the results do not change
significantly which supports the validity of the employed method.}
 \label{fig:borders_comp}
 \end{figure}

As said, the computed boundaries of the C-A plane
are different from what can be found in the literature. Previous
works have been done in the I-band using HST imaging \citep{Ab96,Bri98}. As we have
a sample observed in the I-band, we are able to establish whether the
change in the boundaries has significant consequences in the morphological
classifications. To do so, we classified the I-band sample using the
\cite{Bri98} boundaries and compared the results to the ones obtained with
our method (Fig.~\ref{fig:borders_comp}). We find that there are no significant discrepancies between both classifications. We conclude that our method is valid and moreover has the key advantage being free of subjective judgments.

 \subsubsection{Rest-frame morphologies}
%ZZZ Presenter COSMOS et les images HST ZZ
 We observe some discrepancies in the global morphological distributions between the I and K bands, in particular more perturbed morphologies are seen in the I band.
 When we look at each object individually (Fig.~\ref{fig:morpho_histo_ACS_NACO}) we confirm this trend: there are uncertainties between \emph{K irregulars} and \emph{I disks} and between \emph{I disks} and \emph{K bulges}. Indeed an important fraction of bulge-dominated objects and disk-dominated objects detected by NACO are seen as disk-dominated and irregulars respectively, by ACS, as if there was a systematic trend that moves objects to later types when we move to shorter wavelengths. Certainly, the number of objects is small and a few mismatches cause high discrepancies in Fig.~\ref{fig:morpho_histo_ACS_NACO}. However, this is an expected effect since ACS probes younger stellar populations.  A visual inspection of some of the objects that present different morphologies reveals that some of the ACS irregulars are in fact well-resolved spiral galaxies with inhomogeneities that probably increase the asymmetry indices.  
 
% This distribution is somewhat different from the one measured
%in the K-band as shown in Fig.~\ref{fig:morpho_histo_ACS_NACO}. The
%fraction of irregular objects seems to be more important when measured
%from the I-band whereas bulge dominated objects are more numerous in the K
%band. 

%These differences might come from the fact that we are not observing
%in the same photometric bands and consequently not probing the same
%stellar populations.

 In order to correctly compare both classifications
we need to correct the measurements to estimate how galaxies would look if they were observed locally in the same photometric band. As a matter of fact, \cite{Bri98} showed that high-z galaxies imaged by HST differ in appearance from their local counterparts because of
their reduced apparent size and sampling characteristics, a lower S/N and reduced surface brightness with respect
to the sky background and a shift in the rest wavelength of the
observations. These effects combine to give some uncertainty in the
morphological classifications of galaxies.

The first effect is a change in the concentration value measured at low
redshift. Indeed, \cite{Bri98} draw the boundaries in the C-A plane using a
local sample \citep{Fre96} visually classified. However, the concentration
value depends on redshift, since the threshold is defined relative to
the sky. Thus, less of the galaxy is sampled because of cosmological
dimming. The solution they adopt is to correct C for this effect. We
do not need a correction of the concentration in this paper because  we use
simulations that reproduce exactly the observing conditions to calibrate
the C-A plane. The result is that boundaries are moved with respect to a
local classification instead of changing the C value. 

The shift in the rest-frame wavelength of observations is however
more difficult to estimate. Indeed the question that arises here is whether the morphological type estimated at high redshift
would be the same if observed at low redshift. When observing a galaxy in
the K-band at redshift  $z\sim 1$, the equivalent rest-frame wavelength
will be around the z band, however, when observed in the I-band, the
rest-frame will be around the B band. That implies that a
mismatch can exist in the morphological classification since we are not probing
the same morphological blocks. To correct for this effects we need to
'move the objects' into a common rest-frame wavelength. This is the called
\emph{\emph{morphological k correction}}. The method employed by \cite{Bri98}
consists in determining the morphology from a local sample, redshifting
the objects using SED models, and looking at the fraction of galaxies
that move in to an other morphological class.  A drift coefficient that
characterizes the drift from category X to category Y is thus defined as

\begin{equation}
D_{XY}=\frac{N_{X\rightarrow Y}}{N_X}
\end{equation}
Once the fraction of missclassified objects is determined, the observed number of objects in class X can be related to the true number:
\begin{equation}
N_X^{obs}=N_X+\sum{N_YD_{YX}}-N_X\sum{D_{XY}}
\end{equation}

%In one word if we want to compare our morphology classifications with
%nearby samples we need to correct the morphology. To work in a clean
%way, we should establish the morphology in a local survey and perform
%simulations in order to compute the drift coefficients. However, at the
%moment of writing this paper we do not have this local sample. 

Here we proceed as follows: \cite{Bri98} computed the coefficients to shift
from the I observed morphology to the R rest frame morphology, we use those coefficients to correct the observed HST
morphology of our sample to the one observed in the R rest frame band,
since the filter used for observations is the same (F814W). Once we have this corrected morphology, we can compare it to the NACO uncorrected
morphology. 
%and determine the biases introduced when measuring
%morphology from the K-band and analyzing it in the optical rest frame.
%deduce that way the drift coefficients that must be
%applied to correct from observed K-band to rest frame R band. 
This can
be done because the observed sample is exactly the same in the K and
in the I-band. If we were in the same rest-frame band, we should find the same
morphology.

\begin{figure}
 \resizebox{\hsize}{!}{\includegraphics[scale=.3]{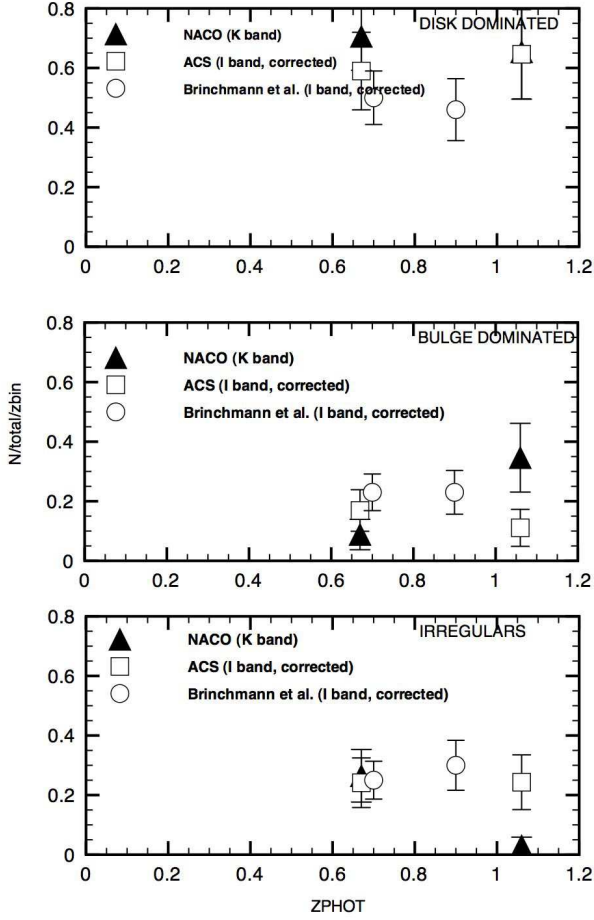}} 
\caption{Redshift distribution for the three morphological types. We
plotted the \cite{Bri98} sample (circles) and our sample observed
with ACS (squares) and with NACO (triangles). \cite{Bri98} and ACS
data are corrected to the R rest-frame band. The NACO sample is
observed from the K-band and no correction has been applied. The ACS
and NACO samples have been separated into two redshift bins ($z<0.8$
and $z>0.8$). The represented redshifts are the median redshifts of each
bin.} 
\label{fig:morpho_evol_comparison} 
\end{figure}

We use the coefficients computed by \cite{Bri98} to correct the ACS
morphology and divide the sample into two redshift bins ($z<0.8$ and
$z>0.8$). Results are shown in Fig.~\ref{fig:morpho_evol_comparison}.

\begin{itemize}

\item At low redshift ($z<0.8$) the corrected I-band distributions and
the uncorrected K-band are similar. The fact that there is no significant difference between the rest-frame R band and the uncorrected K-band population indicates that, when
observing in the K-band, there is no need for \emph{morphological k correction}.
%This implies that there's no
%a significant bias when measuring the R band rest frame morphology from
%the K-band at those redshifts.
\item At higher redshift, the K-band
distribution tends to give a larger fraction of bulge-dominated objects and a lower fraction of irregular objects than the I-band data. This can be caused by the fact that the drift coefficients computed by \cite{Bri98} might contain errors. Since they are based on SED fitting, it is logical to think that errors are more important for high-redshift objects. This would explain why the results are consistent at low redshift. It can also be a consequence of the size of the sample: as the number of objects per bin is small, errors in the classification of a small fraction can lead to discrepancies between the two bands. This issue will certainly be solved with a larger sample.

%Maybe because the disks are too
%faint at those redshifts in the K-band?
\end{itemize}

%which supports the idea that our method is valid.

\begin{table}
\begin{tabular}{c c c c c} 
\hline \hline \noalign{\smallskip} 
\multicolumn{2}{c}{NACO} & & \multicolumn{2}{c}{ACS} \\
morph. type & image & Zphot & image & morph. type\\  
\noalign{\smallskip} \hline \noalign{\smallskip}
Disk dom. & \includegraphics[scale=1.2]{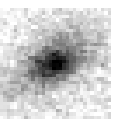} &
1.090 & \includegraphics[scale=1.2]{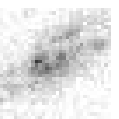} & Irr. \\
\hline \noalign{\smallskip}
Irr. & \includegraphics[scale=1.2]{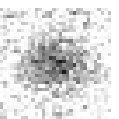} &
0.4223 & \includegraphics[scale=1.2]{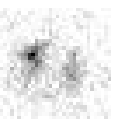} & Irr. \\
\hline \noalign{\smallskip}
Disk dom. & \includegraphics[scale=1.2]{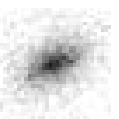} &
0.6689 & \includegraphics[scale=1.2]{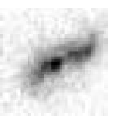} & Irr. \\
\hline \noalign{\smallskip}
Bulge dom. & \includegraphics[scale=1.2]{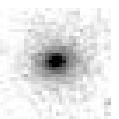} &
1.17460 & \includegraphics[scale=1.2]{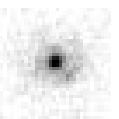} & Disk dom. \\
\noalign{\smallskip}\hline
Disk dom. & \includegraphics[scale=0.6]{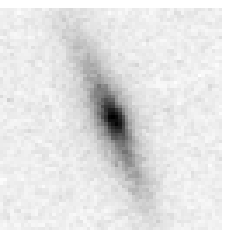} &
0.6689 & \includegraphics[scale=0.6]{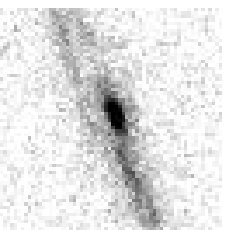} & Irr. \\
\noalign{\smallskip}\hline
Disk dom. & \includegraphics[scale=1.2]{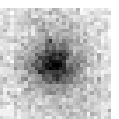} &
0.8861 & \includegraphics[scale=1.2]{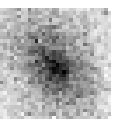} & Irr. \\
\noalign{\smallskip}\hline
Disk dom. & \includegraphics[scale=1.2]{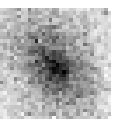} &
0.7394 & \includegraphics[scale=1.2]{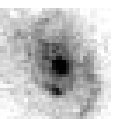} & Irr. \\
\noalign{\smallskip}\hline
\end{tabular}
\caption{\emph{Morphological k correction}: morphological differences when
observing in the K and I-bands. The same objects observed in the K and I bands present different morphologies. The images size is $1.7^{'}\times1.7^{'}$} \label{tbl:ACS_NACO} \end{table}

  %\begin{figure}
 %\centering
 %\includegraphics[scale=0.3]{morpho_evol_ACS.eps}
 %\caption{Morphological evolution of ACS galaxies}
 %\label{fig:morp_evol_ACS}
 %\end{figure}

\section{Summary and conclusions}
We analyzed the morphologies of a sample of $79$ galaxies in the
near-infrared thanks to adaptive optics imaging at a resolution of $0.1^{"}$. Thanks to extensive simulations, we showed that adaptive optics can be used to
obtain reliable high-resolution morphological information in an automated
way and is thus adapted to large observation programmes:
\begin{itemize}
\item Galaxy parameters (bulge fraction, disk scale length, bulge
effective radius) can be estimated by means of model fitting with errors lower than $20\%$ up to $K_s=19$. This is comparable with space data.
\item Galaxies can be separated into three main morphological bins up
to $K_s=21$, with at least $70\%$ of good identifications and $66\%$ up to $K_s=22$.
\item For fainter galaxies ($22<K_s<23$) the accuracy of the morphological classification decreases. Simulations show that bulge-dominated and disk-dominated galaxies can be separated with only $55\%$ accuracy at $K_s=23$.
\end{itemize}

We obtain, for the first time, an estimate of the mix of morphological types of the galaxy population up to $z\simeq1$ from ground-based K-band observations with high spatial resolution comparable or better than visible imaging from space. We demonstrate that estimating morphology from K-band data at $z\simeq1$ is not affected by \emph{morphological k correction}, as there is no significant difference between our population and the corrected I-band population. 
%Despite the still small sample at hands, the galaxy counts are demonstrated to be in good agreement with previous works. 
We find that the fraction of irregulars at $z\simeq1$ is about $12\%\pm2.7\%$ using automated classification methods. This is higher than what is found in local surveys, confirming the well-established trend toward an increasing fraction of irregular galaxies with redshift as observed from surveys in the visible.  Our small sample does not allow us to reach firm conclusions on the evolution of the fraction of late-type or early-type galaxies, but classifying galaxies from K-band AO imaging data is demonstrated to be reliable. 
From this work it seems clear that adaptive optics can be used for
observational cosmology with reliable accuracy, and consequently data of this type should contribute to a better understanding of galaxy evolution in the future. However, it is still a
new technique and technical difficulties exist, such as variable PSF,
small fields, subsampling and the need of guide stars that make the use of
classical reduction methods more difficult. This is now easier with laser guide stars becoming available and new sets of utilities like the ones we are developing to enable easy data processing and analysis of adaptive optics data for the community. This opens the way to observing the large samples required to reach a robust statistical accuracy. We are planning to enlarge our sample by observing a large number of areas around bright stars in the COSMOS field, which will strongly reduce uncertainties in the study
of morphological evolution.

\begin{acknowledgements}

The authors want to thank the anonymous referee for many useful comments that greatly improved this paper. 

\end{acknowledgements}

\bibliographystyle{aa}
\bibliography{biblio.bib}

\clearpage

\begin{longtable}{cccccccccc}
\caption{\label{table} Summary of the morphological classifications for the 79 detected objects. For each object we show the I and K band magnitudes and the estimated morphological type from AO imaging in the K-band (with GIM2D and C-A) and from HST-ACS in the I-band. \textbf{BD stands for bulge-dominated, DD for disk-dominated and I for irregular.}}\\
\hline\hline
Object ID& RA & DEC & Ks & I & rh(arcsec) & ZPHOT & G2D(K) & C-A(K) & C-A(I)\\
\hline
\endfirsthead
\caption{continued.}\\
Object ID& RA & DEC & Ks & I & rh(arcsec) & ZPHOT & G2D(K) & C-A(K) & C-A(I)\\
\hline
\endhead
\hline
\endfoot
\object{NHzG J100016.4+021555} & 150.069 & 2.26541 & 20.1822 & 99.5000 & 0.296730& 99.9000 & DD & DD & N/A\\
\object{NHzG J100016.3+021643} & 150.068 & 2.27872 & 20.7045 & 22.4291 & 0.333504& 0.711700 & DD & DD & I\\
\object{NHzG J100017.2+021637} & 150.072 & 2.27703 & 22.8723 & 24.1406 & 0.205416& 0.247200 & DD & DD & DD\\
\object{NHzG J100016.3+021631} & 150.068 & 2.27552 & 19.7637 & 24.2207 & 0.566514& 1.78470 & DD & DD & I\\
\object{NHzG J100015.0+021629} & 150.063 & 2.27500 & 18.7053 & 99.5000 & 0.489564& 99.9000 & DD & DD & N/A\\
\object{NHzG J100014.7+021630} & 150.061 & 2.27509 & 22.1654 & 23.3978 & 0.209682& 0.832200 & BD & DD & DD\\
\object{NHzG J100014.8+021629} & 150.062 & 2.27487 & 22.0163 & 23.7523 & 0.169614& 0.793700 & I & I & DD\\
\object{NHzG J100014.8+021629} & 150.062 & 2.27478 & 21.3202 & 99.5000 & 0.292356& 99.9000 & I & I & N/A\\
\object{NHzG J100015.9+021629} & 150.067 & 2.27476 & 21.7583 & 22.6455 & 0.246294& 0.0400000 & BD & DD & DD\\
\object{NHzG J100017.2+021624} & 150.072 & 2.27355 & 20.0909 & 23.4449 & 0.243324& 0.956100 & BD & DD & DD\\
\object{NHzG J100017.3+021620} & 150.072 & 2.27223 & 20.2320 & 99.5000 & 0.403920& 99.9000 & I & I & N/A\\
\object{NHzG J100014.8+021624} & 150.062 & 2.27360 & 19.8198 & 99.5000 & 0.231174& 99.9000 & BD & BD & N/A\\
\object{NHzG J100016.7+021618} & 150.070 & 2.27174 & 19.0273 & 20.9238 & 0.640170& 0.219100 & DD & DD & DD\\
\object{NHzG J100016.7+021610} & 150.070 & 2.26963 & 20.3879 & 22.7999 & 0.271026& 0.740000 & BD & DD & BD\\
\object{NHzG J100014.9+021607} & 150.062 & 2.26870 & 18.5753 & 21.8364 & 0.392418& 0.875600 & BD & BD & DD\\
\object{NHzG J100014.9+021603} & 150.062 & 2.26755 & 22.2439 & 22.7490 & 0.235656& 0.475700 & BD & DD & BD\\
\object{NHzG J100017.3+021601} & 150.072 & 2.26712 & 20.0121 & 23.8199 & 0.266760& 1.33790 & DD & DD & I\\
\object{NHzG J100053.2+021934} & 150.222 & 2.32621 & 18.5348 & 21.3553 & 0.531684& 0.886100 & DD & DD & DD\\
\object{NHzG J100051.9+022011} & 150.216 & 2.33658 & 22.6880 & 22.6407 & 0.183600& 0.810100 & BD & DD & DD\\
\object{NHzG J100052.0+022008} & 150.217 & 2.33579 & 19.2141 & 21.9404 & 0.250614& 0.832600 & DD & BD & BD\\
\object{NHzG J100051.8+022006} & 150.216 & 2.33511 & 21.9112 & 23.6353 & 0.254880& 0.818800 & DD & DD & I\\
\object{NHzG J100050.8+022002} & 150.212 & 2.33398 & 20.2005 & 21.8439 & 0.471420& 0.514000 & DD & DD & DD\\
\object{NHzG J100050.8+022000} & 150.212 & 2.33337 & 17.6825 & 19.9774 & 0.690012& 0.400600 & DD & DD & DD\\
\object{NHzG J100051.9+021944} & 150.217 & 2.32890 & 18.7411 & 20.8672 & 0.396252& 0.670000 & DD & DD & DD\\
\object{NHzG J100051.0+021942} & 150.213 & 2.32853 & 18.3509 & 20.8194 & 0.614520& 0.739400 & DD & DD & DD\\
\object{NHzG J100052.4+021941} & 150.219 & 2.32814 & 20.8822 & 22.7814 & 0.404622& 0.799500 & I & I & DD\\
\object{NHzG J100011.1+020629} & 150.047 & 2.10808 & 17.7249 & 19.9124 & 0.348300& 0.506000 & BD & BD & DD\\
\object{NHzG J100010.0+020624} & 150.042 & 2.10673 & 19.5877 & 22.7680 & 0.200826& 0.987200 & DD & BD & DD\\
\object{NHzG J100011.6+020617} & 150.048 & 2.10486 & 19.5207 & 21.0617 & 0.346680& 0.270200 & DD & DD & DD\\
\object{NHzG J100010.2+020612} & 150.043 & 2.10360 & 20.3156 & 99.5000 & 0.155034& 99.9000 & DD & DD & N/A\\
\object{NHzG J100010.6+020612} & 150.045 & 2.10348 & 20.5515 & 99.5000 & 0.167130& 99.9000 & BD & DD & N/A\\
\object{NHzG J100009.7+020610} & 150.041 & 2.10279 & 21.2964 & 99.5000 & 0.178092& 99.9000 & DD & DD & N/A\\
\object{NHzG J095951.4+020514} & 149.964 & 2.08732 & 20.8670 & 21.1830 & 0.379188& 0.671400 & I & I & DD\\
\object{NHzG J095953.9+020507} & 149.975 & 2.08549 & 21.8739 & 23.2931 & 0.170046& 0.703700 & DD & DD & DD\\
\object{NHzG J095952.3+020459} & 149.968 & 2.08319 & 19.9646 & 99.5000 & 0.358668& 99.9000 & DD & DD & N/A\\
\object{NHzG J095952.3+020458} & 149.968 & 2.08286 & 21.7629 & 99.5000 & 0.202284& 99.9000 & BD & DD & N/A\\
\object{NHzG J095952.3+020448} & 149.968 & 2.08006 & 21.3778 & 99.5000 & 0.178254& 99.9000 & BD & DD & N/A\\
\object{NHzG J095953.0+020446} & 149.971 & 2.07970 & 20.2794 & 23.9269 & 0.192024& 0.754600 & DD & DD & I\\
\object{NHzG J095951.5+020444} & 149.965 & 2.07890 & 21.1099 & 23.6728 & 0.160164& 1.37440 & DD & BD & DD\\
\object{NHzG J100013.8+020838} & 150.058 & 2.14397 & 22.4294 & 22.7827 & 0.185220& 0.240300 & DD & BD & DD\\
\object{NHzG J100015.3+020923} & 150.064 & 2.15644 & 19.5222 & 21.1499 & 0.395604& 0.979400 & BD & BD & DD\\
\object{NHzG J100013.9+020920} & 150.058 & 2.15569 & 21.0909 & 99.5000 & 0.189000& 99.9000 & BD & DD & N/A\\
\object{NHzG J100014.0+020919} & 150.058 & 2.15551 & 21.2097 & 23.5849 & 0.247158& 0.994000 & BD & DD & DD\\
\object{NHzG J100014.1+020918} & 150.059 & 2.15517 & 19.5826 & 23.5148 & 0.193482& 1.26360 & BD & BD & DD\\
\object{NHzG J100015.1+020915} & 150.063 & 2.15427 & 23.0568 & 99.5000 & 0.139968& 99.9000 & BD & DD & N/A\\
\object{NHzG J100016.3+020912} & 150.068 & 2.15348 & 21.6753 & 23.0442 & 0.156600& 0.935900 & DD & DD & DD\\
\object{NHzG J100013.9+020909} & 150.058 & 2.15258 & 21.4149 & 22.7981 & 0.306612& 1.08450 & DD & DD & DD\\
\object{NHzG J100013.9+020903} & 150.058 & 2.15105 & 21.4370 & 23.5968 & 0.185112& 1.11800 & BD & DD & DD\\
\object{NHzG J100015.4+020854} & 150.064 & 2.14850 & 20.9069 & 23.9274 & 0.283014& 1.16300 & DD & DD & I\\
\object{NHzG J100016.1+020854} & 150.067 & 2.14840 & 22.1741 & 99.5000 & 0.168156& 99.9000 & BD & DD & N/A\\
\object{NHzG J100003.4+020711} & 150.015 & 2.11984 & 20.5216 & 22.3691 & 0.432864& 0.714900 & DD & DD & DD\\
\object{NHzG J100003.6+020706} & 150.015 & 2.11842 & 19.0519 & 22.4869 & 0.408024& 1.17460 & DD & BD & DD\\
\object{NHzG J100001.7+020704} & 150.007 & 2.11780 & 17.0060 & 18.8229 & 0.862596& 0.338100 & DD & DD & DD\\
\object{NHzG J100002.5+020701} & 150.011 & 2.11709 & 20.4333 & 99.5000 & 0.220644& 99.9000 & DD & DD & N/A\\
\object{NHzG J100000.9+020701} & 150.004 & 2.11697 & 21.7613 & 22.4334 & 0.235710& 0.787500 & I & I & DD\\
\object{NHzG J100003.2+020700} & 150.013 & 2.11682 & 18.8483 & 22.2691 & 0.440532& 1.06140 & DD & DD & I\\
\object{NHzG J100002.6+020659} & 150.011 & 2.11655 & 17.6231 & 99.5000 & 0.397062& 99.9000 & DD & BD & N/A\\
\object{NHzG J100000.9+020652} & 150.004 & 2.11461 & 19.4302 & 20.6107 & 0.372168& 0.698200 & BD & DD & DD\\
\object{NHzG J100002.1+020650} & 150.009 & 2.11392 & 18.8301 & 21.3752 & 0.445932& 0.689900 & DD & DD & DD\\
\object{NHzG J100003.1+020648} & 150.013 & 2.11341 & 20.3555 & 21.9584 & 0.402084& 0.422300 & DD & DD & I\\
\object{NHzG J100003.1+020642} & 150.013 & 2.11180 & 21.3786 & 23.4545 & 0.283662& 0.770500 & I & I & DD\\
\object{NHzG J100002.2+020641} & 150.009 & 2.11151 & 20.8944 & 21.5886 & 0.404406& 0.334700 & I & I & DD\\
\object{NHzG J100002.1+020635} & 150.009 & 2.10981 & 21.4976 & 99.5000 & 0.294678& 99.9000 & DD & DD & N/A\\
\object{NHzG J095956.8+020423} & 149.987 & 2.07308 & 19.8462 & 23.6537 & 0.332424& 1.00700 & DD & DD & DD\\
\object{NHzG J095956.7+020421} & 149.987 & 2.07260 & 22.6499 & 22.7361 & 0.321570& 1.10000 & DD & DD & I\\
\object{NHzG J095955.6+020420} & 149.982 & 2.07236 & 20.8560 & 22.4354 & 0.202986& 0.689900 & DD & BD & BD\\
\object{NHzG J095955.6+020419} & 149.982 & 2.07204 & 21.3234 & 22.7710 & 0.204984& 0.624900 & DD & DD & BD\\
\object{NHzG J095956.4+020416} & 149.985 & 2.07137 & 20.8485 & 22.3323 & 0.404082& 0.0400000 & DD & DD & BD\\
\object{NHzG J095956.4+020412} & 149.985 & 2.07027 & 19.3325 & 99.5000 & 0.389070& 99.9000 & DD & BD & N/A\\
\object{NHzG J095955.8+020412} & 149.983 & 2.07019 & 22.9091 & 23.4010 & 0.194778& 0.525700 & I & I & BD\\
\object{NHzG J095956.8+020406} & 149.987 & 2.06838 & 20.3433 & 23.9523 & 0.259470& 1.54810 & BD & BD & DD\\
\object{NHzG J095955.8+020407} & 149.983 & 2.06874 & 18.9586 & 22.0782 & 0.264924& 1.20850 & DD & BD & DD\\
\object{NHzG J095955.7+020404} & 149.982 & 2.06801 & 20.4066 & 23.8483 & 0.391338& 1.48830 & I & I & DD\\
\object{NHzG J095956.2+020401} & 149.984 & 2.06707 & 21.2254 & 99.5000 & 0.249372& 99.9000 & DD & DD & N/A\\
\object{NHzG J095955.8+020358} & 149.983 & 2.06630 & 21.2308 & 23.8229 & 0.434916& 0.881200 & DD & DD & I\\
\object{NHzG J095956.5+020355} & 149.986 & 2.06550 & 18.0002 & 21.4034 & 0.622458& 0.794100 & DD & DD & DD\\
\object{NHzG J095956.2+020353} & 149.984 & 2.06479 & 19.3081 & 21.7327 & 0.394146& 0.668900 & DD & DD & DD\\
\object{NHzG J095956.4+020353} & 149.985 & 2.06474 & 22.1424 & 23.8900 & 0.188568& 0.743900 & DD & DD & DD\\
\object{NHzG J095957.2+020347} & 149.989 & 2.06312 & 21.8247 & 23.8017 & 0.184248& 0.214600 & DD & DD & DD\\
\end{longtable}

\clearpage

\end{document}